# Sentiment Analysis in Digital Spaces: An Overview of Reviews

Sentiment Analysis in Digital Spaces

Laura, E. M., Äyräväinen*

University of Bath, lema21@bath.ac.uk

Joanne, Hinds

University of Bath, jh945@bath.ac.uk

Brittany, I., Davidson

University of Bath, bid23@bath.ac.uk

Digital data generated via social media have become a prosperous entity for sentiment analysis (SA) researchers seeking to understand individuals' feelings, attitudes, and emotions. Numerous systematic reviews have sought to synthesize the diverse contexts, media, and methodologies studied across a variety of applications. While these reviews are useful for synthesizing *what* work has been conducted, they rarely address questions about validity of SA methods or critical assessment of scientific practices. Our overview of 38 reviews, comprises 2,275 primary studies using SA to dissect online digital data. We provide a high-level overview of current applications, methods, outcomes, and common challenges in SA research. A bespoke quality assessment framework was designed to assess the rigor and quality of systematic review methodologies and reporting standards. We found diverse applications, methods, outcomes, and persistent challenges outlined in the reviews, which we discuss in relation to validity of SA research. Importantly, the methodological rigor of the reviews was limited, which we therefore discuss considering how existing systematic reviews might influence our understanding of the field and impact the subsequent decisions that researchers and practitioners may take. We discuss how future research can address these issues and highlight their importance across numerous societal applications.

## 1 INTRODUCTION

The proliferation of the internet, social media, and digital devices have resulted in the generation of vast amounts of data. For example, language, or textual data published across websites, online communities, news sites, and social media platforms can reflect individuals' subjective states, such as emotions, opinions, and attitudes (e.g., Giachanou & Crestani, 2016; Qiu et al. 2010; Xu et al. 2012). Since the early 2000's the popularity of sentiment analysis (SA) has grown substantially; where researchers can access sentiment data from diverse populations (i.e., individuals that may be difficult to access otherwise and from countries around the world) at scale. Such data can offer vast insights into feelings about, and public opinions on the societal issues of our time. For example, investigations into people's sentiments during infectious disease outbreaks, epidemics, and pandemics including Ebola and Covid-19 have contributed to a deeper understanding of people's sentiments toward unfolding events, the spread of misinformation, and the prevalence of fake news (e.g., Bangyal et al. 2021; Iwendi et al. 2022). As other examples, research of student feedback in massive open online courses (MOOCS) has been used to leverage insights into dropout rates and influence policies in higher education (e.g., Crossley et al. 2016; Kastrati et al. 2021) and investigations into emotions derived from social media posts have been used to predict behavior on the stock market (e.g., Bollen et al. 2011; Pagolu et al. 2016).

---

* Place the footnote text for the author (if applicable) here.

The rapid rise of SA applications has prompted multiple systematic reviews seeking to synthesize existing work in particular domains. Such reviews have covered topics including online reviews and public opinion (Jain et al. 2021; Kumar & Sharma, 2017; Skoric et al. 2020), Covid-19 (e.g., Aljedaani et al., 2022; Alamoodi et el., 2021), and SA in different languages (e.g., Handayani et al. 2018; Ghallab et al. 2020; Obiedat et al. 2021; Osorio Angel et al. 2021). A common approach applied involves reviewing what methods have been employed in previous research, for instance those used for data collection, data processing, and the models used for prediction (e.g., Asghar et al. 2014; Yadav & Vishwakarma, 2020; Zad et al. 2021). While such approaches are often useful for summarizing methods and for creating taxonomies of the findings across specific applications or sub-disciplines of SA (e.g., Alamoodi et al. 2021; Ghallab et al. 2020), they rarely provide insights into or critical assessment of the validity of different SA methods, such as inspection of how the construct of interest ('sentiment') is defined and operationalized in extant literature, or the extent to which proposed SA systems can be used beyond their original target population, context, or time period. Further, the guidance on performing and reporting systematic reviews in computer science is limited; researchers often consult guidelines provided from medicine (such as the Cochrane Collaboration, e.g., Chalmers, 1993, the EQUATOR network, https://www.equator-network.org/) or software engineering (Kitchenham, 2004; Kitchenham & Charters, 2007), which may not be entirely appropriate or applicable to SA contexts. Consequently, research may be inadequately synthesized, analyzed, and reported, which could weaken the evidence presented, impacting subsequent research decisions and directions. Insights could also misdirect policy and implementations of SA-based tools if synthesized findings are unreliable.

### 1.1 Sentiment analysis in digital spaces: the need for an overview of reviews

Digital spaces encompass a vast range of media from which researchers aim to infer individuals' emotions and attitudes, via social media, blogs, online communities, and news sites. Each of these platforms introduces unique contextual nuances; for instance, the way users communicate and present themselves on Instagram may greatly differ from their interactions on TikTok (Davidson & Joinson, 2021). These disparities are shaped by community norms, established conventions (e.g., De Souza & Preece 2004; Preece & Shneiderman, 2009), and design affordances that influence users' interactions (e.g., Chen et al. 2019; Norman, 1998; Zhao et al. 2013). For example, users of online health communities may be motivated to gain reputation points, prompting them to write informative and supportive posts (e.g., Chen et al. 2019), while users on X (formerly known as Twitter) may seek to generate viral content by creating highly contentious messages (e.g., Pressgrove et al. 2018). The digital landscape also facilitates multimodal communication, where SA can span diverse sources such as conversational/audio data from videos, alongside textual data, likes, emojis and so forth (e.g., Abdu et al. 2021; Perdana & Pinandito, 2018; Shiha & Ayvaz, 2017). Indeed, these distinctive contexts and data modalities diverge significantly from offline or alternative forms of media such as essays (Rude et al. 2004), stories (Alm et al. 2005), and conversations (Huddar et al. 2021; Shenoy & Sardana, 2020), which comprise different styles of writing or speech and tend to focus on language only. The characteristics of online content make it a rich source of data for SA, while also posing specific challenges, such as variation in language use (e.g., misspellings, slang). As such, our overview targets online content, due to its popularity, potential, and persisting challenges for SA.

Although the majority of SA research tends to revolve around discussions embedded within computer science communities, the study of language and emotions in psychology (for example see the work by Pennebaker and colleagues; Chung & Pennebaker, 2011; Pennebaker et al. 2003; Tausczik & Pennebaker, 2010) could provide valuable insights into the way in which researchers conceptualize and operationalize individuals' sentiments in their research. As such, we explore how researchers define sentiment in systematic reviews to investigate (and search for evidence on) particular topics because SA has been applied widely across applied disciplines (e.g., politics), and increasingly SA is being used as a tool within mental health diagnostics (e.g., Rajput, 2020; Wang et al., 2020; Yeow & Chua, 2022), which suffer with their own inconsistences of definitions and therefore measurement of mental health symptomologies (Fried, 2017). Hence, definitions are critical for research because they lay the foundations for scientific inquiry and are influenced by prior evidence or theory or even cultural differences (Fried,2017; Lim 2016)—this is critical in clinical settings or when working with vulnerable people.

Thus, without formal definitions and consensus regarding how we understand a construct, such as sentiment, the ways to measure it across contexts (e.g., verbal cues, body language, in text, audio, video) will likely be inconsistent and incomparable. Further, a lack of 'ground truth' for sentiment (or emotion or feeling more generally) is an example of issues related to a latent construct, wherein lies difficulty in measuring something inherently unobservable. Therefore, our conceptualizations, definitions, and measurement of sentiment rely on natural language (akin to much of psychological constructs), meaning that our consistency of measurement remains questionable and thus will have a knock-on impact on reproducing, replicating, and generalizing from current work (Bollen, 2002; Yarkoni, 2020).

Systematic reviews are often considered to be the most valid approaches for evaluating research findings, according to the so-called "hierarchy of evidence" that appraises different methodologies in terms of their effectiveness, appropriateness, and feasibility (e.g., Evans, 2003; Kitchenham, 2004). By providing a "complete" picture of evidence on a given topic, readers can evaluate the strengths and weaknesses of existing research to make informed decisions about future actions (MacKenzie et al., 2012). Systematic reviews can therefore be highly informative for the study of SA in digital spaces because they can educate readers on the most effective techniques and strategies for analysis and can identify gaps in current findings, guiding future researchers to address them. This is especially pertinent given the many varied data modalities and platforms that evolve rapidly amidst technological advances and unfolding events. Thus, systematic reviews can provide cutting-edge insights into public opinion and sentiments surrounding societal and global issues. They can inform policy makers' decisions and responses to public events, such as political developments, natural disasters, the spread of misinformation and hate speech. Systematic reviews can also direct technical decisions, designs, and implementations based on a collection of evidence that should provide direction on what has worked well and what has been less successful, thus providing evidence to help research focus on fruitful avenues of work. An overview of reviews can therefore synthesize a broad array of evidence and identify key trends, gaps and persistent challenges across an entire field that could not be identified otherwise. We therefore investigate the topics, methods, findings, and challenges identified in systematic reviews of SA in digital spaces.

It is critical that systematic reviews are performed meticulously and comprehensively in order to effectively use their findings in evidence-based decision making and in subsequent research (note - this also applies to the insights derived from an overview of reviews). In the fields of medicine and healthcare (from where systematic reviews originated, Cochrane, 1972; Chalmers, 1993) it is widely expected that researchers will adhere to strict protocols when performing systematic reviews (MacKenzie, 2012; Moher et al. 2015; Page et al. 2021). However, fields outside of medicine often have no (or limited) discipline-specific guidelines for conducting systematic reviews, creating situations where researchers commonly consult guidelines from medicine and adjust them to suit their project needs. Moreover, researchers performing systematic reviews of SA often possess expertise in SA or computer science methodologies rather than systematic reviewing techniques. This stands in contrast to medical researchers who are typically specialists in systematic reviews. Consequently, SA researchers might lack the experience and expertise required to formulate new protocols tailored toward reviewing SA in digital spaces. For instance, a researcher conducting a risk of bias assessment may attempt to deconstruct the reporting of methodological procedures such as the model set up, tuning processes, data processing and evaluation to investigate the factors that may influence the validity of the findings, and whether publication bias may exist. However, they may not have systematic reviewing expertise to formulate and apply such criteria in this context. Similarly, SA experts may not have the expertise, awareness, or motivation to document their methods and findings in a way that supports the detail and transparency required to perform a systematic review. Researchers conducting systematic reviews may therefore misinterpret the methods and results or incorrectly apply systematic reviewing protocols to existing work. As a result, the quality and rigor of the procedures applied may be compromised, leading to inconsistent interpretation and application of existing frameworks. Alternatively, some researchers might even resort to devising their own reviewing protocols, or in some cases overlook the application of protocols entirely.

Adhering to procedures that are well-defined and structured ensures that the review is systematic, as well as replicable and free from bias (e.g., Page & Moher, 2017; Sarkis-Onofre et al. 2021). Effective systematic reviews are therefore reliant on the detailed and thorough reporting of the research, from the conceptualization of the research questions, through to the findings and conclusions. Concerns over the inaccurate, misleading, or incomplete reporting of machine learning models are being increasingly reported in fields such as medicine (e.g., Christodoulou et al. 2019; DeMassi et al. 2017; Faes et al. 2020; Yusuf et al. 2020). These have been reflected in increasing calls for computer scientists to document their methods transparently, for instance see "model cards for model reporting" (a framework for documenting the detailed performance of model characteristics, Mitchell et al. 2019) and "datasheets for datasets" (a framework for the documenting the provenance, creation, and use of machine learning datasets to avoid discriminatory outcomes, Gebru et al. 2021) amidst changes in journal policies to promote computational reproducibility (the sharing of data and code to support the replication of the published findings, Stodden et al. 2018). However, we do not currently know the extent to which systematic reviewing procedures are followed, or how transparently methods and findings are reported within the context of SA in digital spaces.

This article therefore aims to address this gap by presenting an overview of reviews (i.e., a review of systematic reviews) that synthesizes existing systematic reviews. As such, we first investigate the current state-of-the-art of research by collating the diverse range of topics, methods, findings, and challenges of SA research. We then develop a bespoke quality assessment framework designed to assess the rigor and quality of systematic review methodologies and reporting standards of existing research. This evaluation is further complemented with inspection of how the construct of interest ('sentiment') is defined and the extent to which open science practices are followed in the current research. By evaluating the quality of research, we discuss the strength of existing findings and identify directions for future research. Our findings provide a comprehensive resource that can aid researchers, practitioners, data scientists, and policy makers in making decisions surrounding research design and the application of SA in digital spaces in future work.

## 2 METHOD

An overview of reviews is a synthesis of results from several systematic reviews and meta-analyses on a particular topic. The purpose of an overview of reviews is to summarize and evaluate research evidence at a broad level, and to provide an “entry point” for readers to access more detailed findings reported in the systematic reviews and primary studies (Caird et al. 2015; Hunt et al. 2018). Accordingly, to synthesize existing systematic reviews, we followed the Preferred Reporting of Items for Overviews of Reviews (PRIOR) (Gates et al. 2022), which is akin to the PRISMA protocol (Page et al., 2021). The PRIOR checklist outlining where the present overview meets each of the criteria is available on the Open Science Framework (OSF, here), along with all supplementary materials associated with this project.

**Protocol and Pre-registration.** The research protocol was pre-registered before data extraction and analysis (available here). During our research, we found it necessary to update and make a. number of deviations from our original pre-registered protocol. These modifications were essential to extract more detailed and suitable information from the reviews. A detailed overview of these deviations are listed in a separate document on OSF (file Protocol_Amendments.xlsx, here).

**Eligibility criteria.** Only systematic reviews, meta-analyses or mapping studies were included, which we defined as studies that report a reproducible search strategy implemented in at least one database, including year range, clearly outlined search terms and inclusion/exclusion criteria for the primary studies. Additionally, the inclusion criteria for the reviews were that they (i) evaluate, summarize or compare SA tools, (ii) include primary studies that use digital data (e.g., from online platforms or devices) and (iii) include empirical studies. During our full-text eligibility assessment, we enforced an additional requirement: the study’s search terms had to align with one of our search terms for sentiment analysis (see below in Search Strategy) or a synonym of these terms. This criterion was included to ensure that we capture only studies that aimed to review SA methods, rather than studies with a broader focus on natural language processing (NLP) or social media analytics. All studies reported in English and available on the final date of search (31st of January 2023) were included. The eligibility of articles was not restricted by publication status; both published and unpublished documents were included (e.g., peer-reviewed journal articles, preprints, and dissertations). Studies examining any human populations, regardless of age, gender, or geographic location were included.

**Information sources.** We consulted four electronic databases in performing our search. These included: ACM Digital Library, IEEE Xplore, PsychInfo (including PsychExtra for unpublished papers) and Scopus. This combination of databases enabled us to perform an exhaustive search, and to capture a breadth of literature on SA across the computational (ACM Digital Library and IEEE Xplore) and psychological (PsychInfo and PsychExtra) sciences as well as across other disciplines that may have conducted such work (Scopus).

**Search Strategy.** Our first step was to search for all overviews of reviews of SA across all online domains and to note the general topics and findings reported. This was to ensure that we did not duplicate existing overviews of reviews that had already been conducted. Our findings here informed any changes to our pre-registration and allowed us to focus more specifically on certain aspects of SA. This process resulted in the identification of one previous overview of reviews of SA by Ligthart et al. (2021), which focused on systematic reviews of text-based SA. Our overview differs from theirs in that (1) we synthesized only systematic reviews conducting SA with online data, (2) we did not restrict data modality to text-based SA and (3) we focused on the methodological and reporting quality of the reviews by developing a quality assessment framework.

The final search string was arrived at after pilot-testing several search string variations. The aim of this exercise was to capture as many relevant documents as possible, while excluding documents that were irrelevant for our research aims. For instance, the preliminary searches revealed a significant number of survey papers. Browsing through several of these documents led us to exclude ‘survey’ as one of our search terms because these articles tended to be non-systematic in nature (e.g., narrative reviews, editorials, or commentaries). This exercise also revealed that several studies used the terms "sentiment analysis" and "opinion mining" interchangeably. Thus, we included both terms in our search (see section 4.1.1 for discussion). After iterative testing of different search term combinations, our final search string consisted of three clauses of search terms, which focused on sentiment analysis, online contexts, and types of reviews, as follows:

"sentiment analysis" OR "sentiment detection" OR "sentiment classification" OR "opinion mining" AND online OR platform OR community OR forum OR "social network*" OR "social media" OR facebook OR twitter AND SLR OR "Systematic review" OR "systematic literature review" OR "systematic mapping" OR "mapping study" OR "meta analysis" OR "meta-analysis" OR "taxonomy" OR "scoping review"

Papers with titles or abstracts that contained at least one term listed in each search clause were included (e.g., a paper including "sentiment analysis", "social network" and "systematic review" in its title or abstract). Finally, we conducted a further search for systematic reviews by extracting the citations of systematic reviews analyzed in other overviews of reviews. This resulted in searching through the systematic reviews studied in the one overview study identified (i.e., by Ligthart et al., 2021).

### 2.1 Selection process

The documents identified via the literature search were imported into Zotero and duplicates were removed. The documents were then screened for inclusion, first based on titles and abstracts, and then based on the full text of the documents. Figure 1 depicts a PRISMA flowchart of the study selection process.

**Document inclusion.** In the title-abstract phase, the first author (LA) screened all 210 documents. A sample of 82 documents were then screened by the other two reviewers (second and third authors, henceforth JH and BID), 41 unique documents each. Half of these 82 documents were categorized as 'maybe' by LA and were used as a basis for the calibration discussions. The remaining 41 documents were used to assess inter-rater agreement. Cohen's $\kappa =$ 0.72 for 21 documents screened by LA and JH and 0.6 for 20 documents screened by LA and BID indicated a moderate and substantial agreement between raters, respectively (Landis & Koch, 1977). Discrepancies were discussed and addressed by the reviewer team, after which the remaining documents were screened again by LA.

In the full text phase, 94 documents (including additional five documents from Ligthart et al., 2021, see Figure 1) were screened by LA. A sample of 54 documents were then screened by JH and BID, where again 20 documents marked as 'maybe' by LA were used as a calibration exercise (5 screened by all three reviewers, 8 by LA and JH and 7 by LA and BID) and the remaining 34 documents were used to assess inter-rater agreement. Cohen's $\kappa$ = 0.88 for 17 documents screened by LA and JH and 0.76 for 17 documents screened by LA and BID indicated substantial agreement between raters. Discrepancies from the screening were discussed by the reviewer team before the rest of the documents were screened again by LA. 41 documents were included in the data extraction phase, but three of these documents were later deemed not eligible during the data extraction process, after a discussion with LA, JH, and BID. This was because more thorough examination of these documents revealed that two of them were not strictly systematic based on our criteria and one of them did not focus on SA. Hence, we then proceeded with 38 documents. Figure 2 displays the screening process and the full list of excluded studies with reasons for exclusion are available on OSF (file Full_Text_Exclusion.xlsx, here).

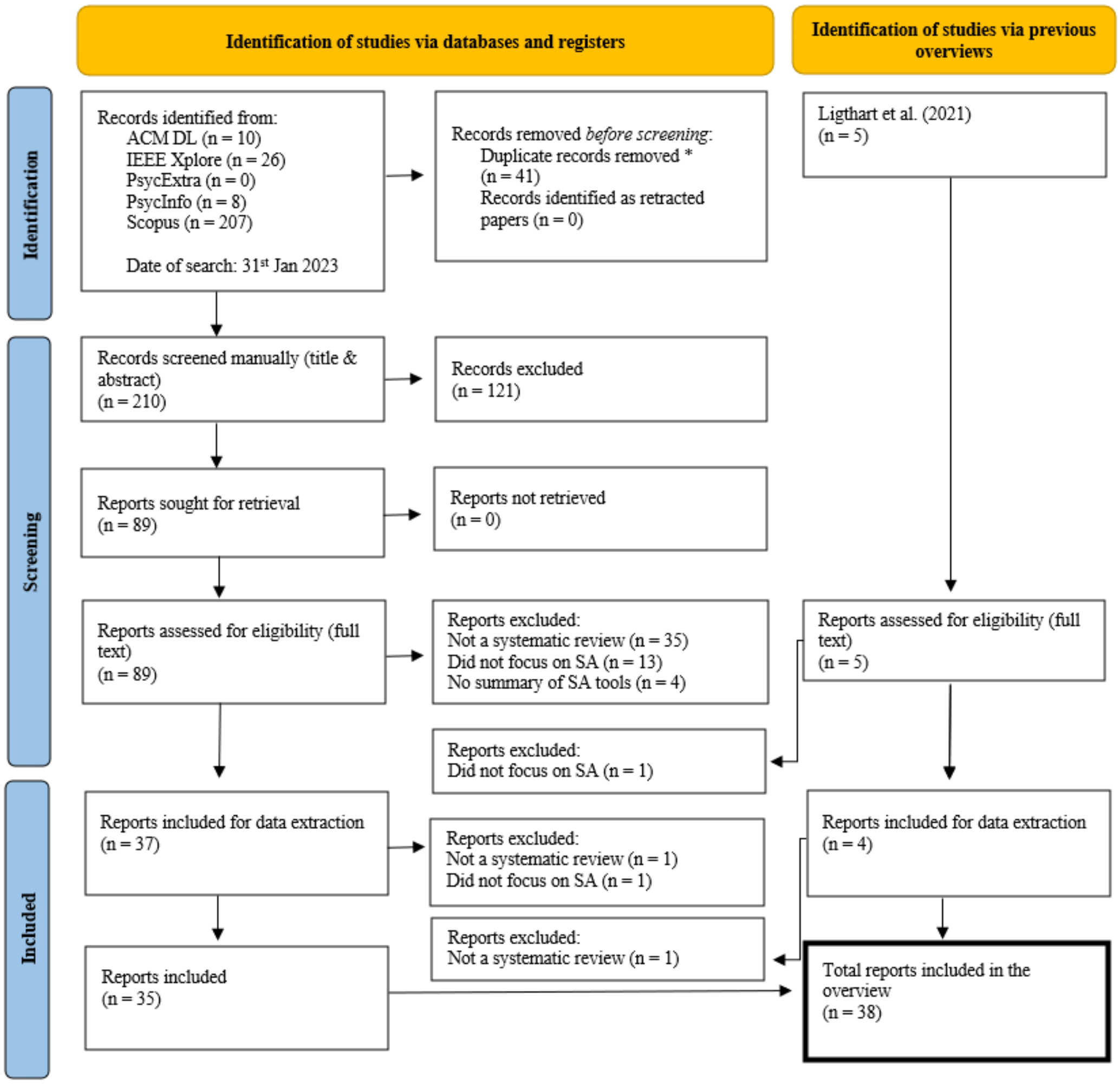

*Records automatically identified as duplicates in Zotero were checked manually before removal

Figure 1: PRISMA flow chart of study selection

**Overlap of reviews.** The list of primary studies included in each of the reviews were extracted and compared by LA. Overall and pairwise overlap percentages, estimated as corrected covered area (CCA) were then calculated using the ccaR package (Bougioukas et al. 2023) on R (version 4.2.0). The overlap of primary studies was assessed for all the primary studies that could be confirmed from the reviews. A total of 127 primary studies were missing due to insufficient reporting in the reviews, including all of the primary studies for one review. Thus, 37 reviews were included in this assessment. The analysis consisted of 2,148 primary studies, of which 1,935 were unique. The primary studies across reviews were matched based on first author, year and title. The form of the citations was unified by removing special characters, typographical errors and by capitalizing the author-year-title strings.

### 2.2 Extraction and Coding of Study Data

**Data collection process.** The data from the studies were extracted by LA, and a sample of 10 reviews was validated by JH and BID. In the event of discrepant data for the same primary study presented in different reviews, we consulted the original primary study for each review (e.g., the version of the primary study used for data extraction), and if necessary, contacted the authors of the reviews reporting discrepant data.

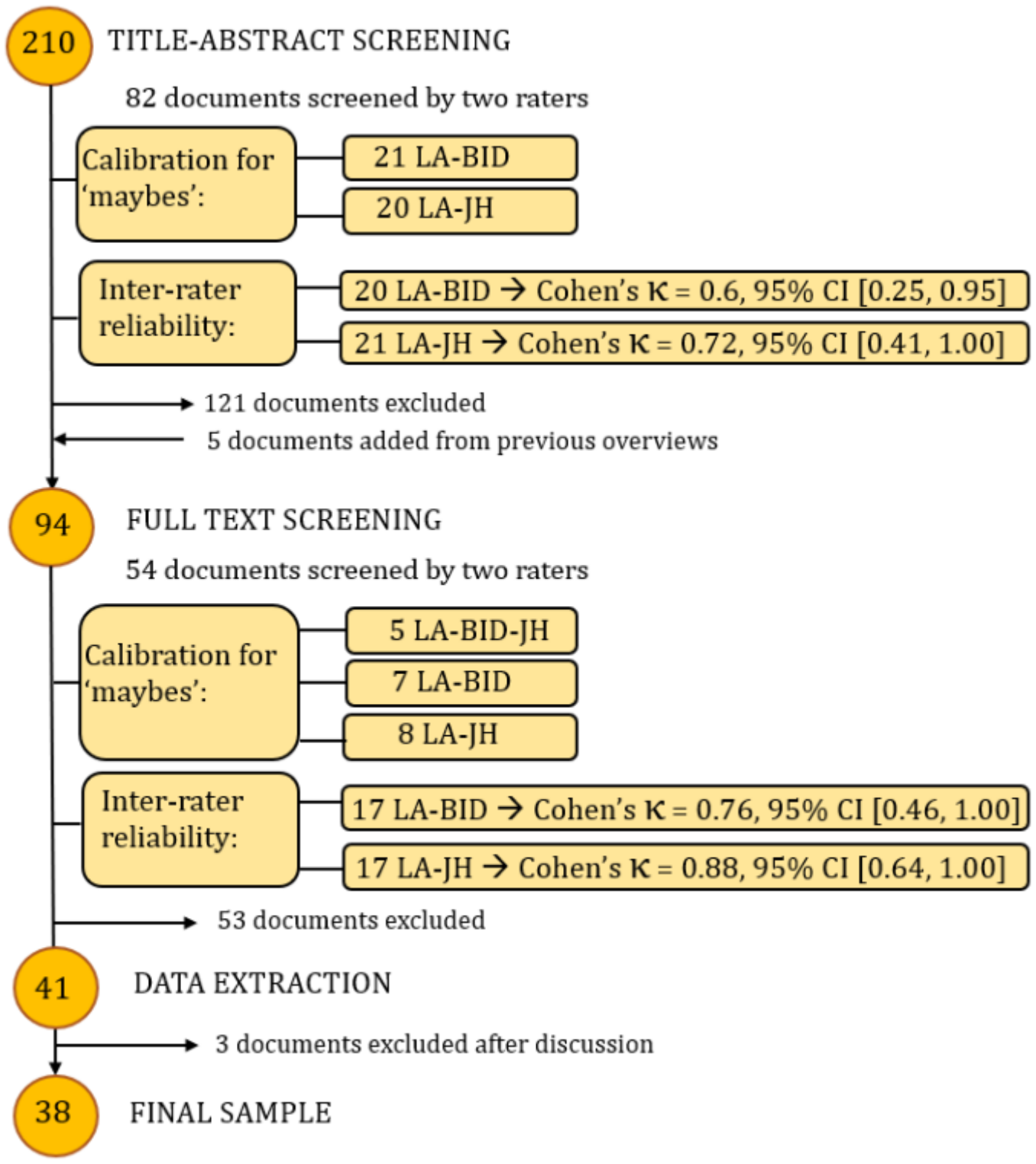

Figure 2: The process of document screening

**Data items.** The following basic descriptors of the records were extracted: Authors, year, title of the study, publication status, journal name (for published articles) and DOI. Additionally, the following characteristics were extracted from the reviews to summarize the key findings regarding SA methods, applications and challenges discussed in the reviews:

- Data source, modality, and language used in the primary studies (Table 3)
- Pre-processing steps and features extracted in the primary studies (Table 4)
- Sentiment classifiers and lexicons used in the primary studies (Table 4)
- Applications and outcomes of SA in the primary studies (Table 5)
- Challenges discussed in the reviews (section 3.3)

All data extraction was performed by LA. Extracted data regarding SA methods, applications and outcomes were then summarized and tabulated. For the extracts about challenges discussed in the reviews, the discussion points were first inspected and grouped into themes before tabulation.

### 2.3 Quality assessment framework

To assess the methodological quality of the reviews, we developed a framework that assessed the design, method, implementation, and reporting presented in each article. The need to improve reporting and data sharing has been widely discussed in fields including medicine and psychology (for example see Patzer et al., 2021; Roche et al., 2015; Towse et al., 2021; Yusuf et al., 2020). However, related discussions or recommendations for reporting research are not prevalent within computer science communities (to our knowledge). Similarly, there are no guidelines/frameworks for assessing the risk of bias or the methodological quality of computer science reviews. We therefore consulted a range of frameworks and checklists from different disciplines and applied/adapted relevant criteria to the present study. These included Hinds et al.'s (2021) framework that assesses reporting standards and the practical utility of automated approaches to predicting personality, as well as the PRISMA (Page et al., 2020) and PRIOR (Gates et al. 2022) guidelines, which delineate protocols for reporting systematic and overviews of reviews respectively. For each criterion, we extracted associated data and developed a scoring system that assessed whether each criterion was reported (yes/no, or yes/partly/no – Table 1 provides a breakdown of the scoring system applied). To explore the influence of review guidelines on the quality of the reviews, we also compared the quality of reviews based on whether and which review guidelines were used in each review.

### 2.4 Open science, reproducibility, and replicability

We further inspected how the term 'sentiment' was defined in the reviews, to gauge the level of consensus about this core concept of the field, as unclear definitions may lead to inconsistent operationalization of constructs of interest, thus hindering replicability of research. This involved extracting any definitions provided in the reviews and grouping this information based on whether a definition was provided, and how it was defined (i.e., which terms were used to define sentiment). We also tracked the accessibility of the reviews (i.e., whether the reviews were open access or otherwise freely available) and any related materials (e.g., whether any protocols or code were shared), as additional indicators of open science practices in SA research. This is because by nature systematic reviews (and any systematic methods such as meta-analyses) should be open and reproducible. This is critically important to ensure the way in which the project was composed and executed is credible and verifiable. Hence, we deemed it important to capture how open, reproducible, and replicable the reviews are. The full list of extracted data is presented in supplementary materials on OSF (file Data_Extraction.xlsx, here).

Table 1: Scoring criteria of the quality assessment framework

| Quality criterion | Yes | Partly | No |
|---|---|---|---|
| Guidelines | The systematic review is informed by review guidelines (such as those written by Kitchenham, 2004, Tranfield et al., 2003, or the PRISMA framework, Page et al., 2020). | The systematic review cites other reviews as inspiration and/or uses the PICOC framework to formulate a research question but does not use any specific guidelines. | The systematic review cites no other sources or guidelines in performing the review whatsoever. |
| Rationale | A clear and well supported rationale for conducting the systematic review is provided. | A rationale for performing the systematic review is provided but is limited in terms of the importance of the topic or need for a systematic review. | A clear justification for conducting the systematic review is missing. |
| Objective | A clear list of research questions/aims is provided. | A clear list of research questions is missing but the general aim of the review is expressed | The aim of the review is unclear |
| Inclusion/Exclusion | Inclusion/exclusion criteria for the systematic review are specified. | - | Inclusion/exclusion criteria for the systematic review are not specified. |
| Search Strategy: Study Identification | An exhaustive search strategy to identify studies is reported. This includes the adoption of multiple approaches, which may include (some combination of) searching multiple databases, forward/backward searching, announcements (making calls for papers), and manual searching. | A search strategy is employed but may not be exhaustive based on the detail reported. For instance, a systematic review may report a single database search with no further searches. | A search strategy is not reported. |
| Search Strategy: Search Terms | A clear and exhaustive list of search terms is reported, including specification of the relationship of the terms (e.g., use of Boolean operators). | Search terms are clearly reported but their relationship (e.g., Boolean operators) is unclear. | A complete list of search terms is missing from the systematic review. |
| Search Strategy: Search dates | The date of the search is reported and includes the month and year that the search was performed. | The date of the search is reported and includes the year that the search was performed. | The date of the search is not reported. |

| Quality criterion | Yes | Partly | No |
|---|---|---|---|
| Included Studies | All the primary studies are identifiable from information provided in the systematic review. For example, this could include tables that present information on each study, or reference lists provided in supplementary materials. | All the primary studies are identifiable but only by inferring information provided in the article. For example, this could include a table reporting some of the study characteristics, or from a summary describing aspects of the studies included. | Identifying all the primary studies is not possible from the information provided. |
| Validity of Screening: Methods | A method (or multiple methods) to check for/assess the validity of screening is reported. These methods could include calibration exercises, voting, screening, or equivalent procedures performed by multiple reviewers. Clear information about the number of studies included, the number of reviewers involved, and the way discrepancies were handled are reported. | A method (or multiple methods) to check for/assess the validity of screening are reported. These methods could include calibration exercises, voting, screening, or equivalent procedures performed by multiple reviewers. However, information about the number of studies included, the number of reviewers involved, and the way discrepancies were handled is incomplete or missing. | Information about adopting methods to ensure the validity of the screening process is insufficient or missing. |
| Validity of Screening: Inter-rater Reliability | Interrater reliabilities are reported. | - | Interrater reliabilities are not reported. |
| Validity of Data Extraction | A method (or multiple methods) to check for/assess the validity of data extraction from the primary studies is reported. These methods could include calibration exercises, voting, data extraction, or equivalent procedures performed by multiple reviewers. Clear information about the number of studies included, the number of reviewers involved, and the way discrepancies were handled are reported. | A method (or multiple methods) to check for/assess the validity of data extraction are reported. These methods could include calibration exercises, voting, data extraction, or equivalent procedures performed by multiple reviewers. However, information about the number of studies included, the number of reviewers involved, and the way discrepancies were handled is incomplete or missing. | Information about adopting methods to ensure the validity of the data extraction process is insufficient or missing. |
| Quality Assessment of the Primary Studies | A valid assessment of the quality of the primary studies is reported using an approach designed by the researchers or a pre-existing framework. | An assessment of the quality of the primary studies is performed using an approach designed by the researchers, but the approach may be limited or invalid (for example using journal selection as a means to assess quality). Alternatively, a reasonable justification was provided for not performing a quality assessment. | An assessment of the quality of the primary studies is not reported. |
| Limitations | A discussion of the limitations of the systematic review is reported. | - | A discussion of the limitations of the systematic review is not reported. |
| Sampling | Whether there was any discussion regarding sampling in the review (yes/no). | - | No discussion regarding sampling was present in the review |
| Pre-registration | The systematic review reports details of pre-registration on a relevant website such as PROSPERO, or OSF. | - | The systematic review reports no details of pre-registration. |

# 3 RESULTS

## 3.1 Characteristics of the reviews

Our final sample comprised 38 articles in total, which included 33 systematic reviews, two scoping reviews, two systematic mapping studies and one meta-analysis (we refer to these collectively as reviews). The key information for each review is provided in Table 2 and the full reference list of each included review is provided in Appendix A.

As seen in Table 2, the majority of the 38 reviews were published in 2020 or later; 70% of these were published between 2020-2022, and the earliest review was published in 2013. The temporal coverage of the reviews ranged from two to 13 years, and the year 2013 was most often covered in different reviews (10 reviews). Based on the subject areas listed for each journal in Scopus, the most common disciplines in which the reviews were published (including multidisciplinary journals/conferences) were computer science (85%) and engineering (40%). Five reviews were excluded from these calculations (Agustiningsih et al. 2021; Elnagar et al. 2021; Handayani et al. 2018; Pinto et al. 2021; Tho et al. 2020), as these were published in conferences for which the subject area was not listed in Scopus.

The overlap of primary studies across the reviews was estimated as overall and pairwise CCA. The overall CCA across the 37 reviews was 0.3%, and is considered slight following Pieper et al. (2014) thresholds. Four review pairs shared a moderate overlap (6-10%) and one review pair reached an overlap of 28.7%, considered very high. All the review pairs with moderate or higher overlap were in the health domain. See Figure 3 for a heatmap depicting each pairwise CCA.

Table 2: Key information of the reviews

| Study | Study ID | Primary studies | Year range of included studies | Review type | Article type | Discipline | Inclusion |
|---|---|---|---|---|---|---|---|
| Abdullah & Rusli 2021 | S1 | 45 | 2011-2018 | SLR | Journal | AgriBioSci; ChemEng; CompSci; EnvirSci | Studies using multilingual SA, published 2010-2019 |
| Agustiningsih et al. 2021 | S2 | 21 | 2020-2021 | SLR | Conference | - | Studies using SA on Covid-19 vaccine posts in Twitter, published 2020-2021 |
| Ahmad et al. 2018 | S3 | 8 | 2013-2016 | SLR | Journal | CompSci | Studies using SVM for SA, published 2012-2017 |
| Alamoodi et al. 2021a | S4 | 33 | 2017-2021 | SLR | Journal | CompSci; Med | Studies using SA on vaccine hesitancy, published 2010-2021 |
| Alamoodi et al. 2021b | S5 | 28 | 2013-2020 | SLR | Journal | CompSci; Eng | Studies using SA on infectious diseases, published 2010-2020 |
| Aljedaani et al. 2022 | S6 | 47 | 2020-2022 | SLR | Journal | ComSci; Eng; Math | Studies using AI techniques for SA about Covid-19 vaccine, published 2020-2021 |
| Al-Moslmi et al. 2017 | S7 | 28 | 2010-2016 | SLR | Journal | ComSci; Eng; MatSci | Studies using cross-domain SA, published 2010-2016 |
| Cortis & Davis 2021 | S8 | 485 | 2010-2018 | SLR | Journal | A&H; CompSci; SocSci | Studies using opinion mining or SA on social media data, published 2007-2018 |
| Dalipi et al. 2021 | S9 | 40 | 2014-2021 | SLR | Journal | CompSci | Studies using SA on MOOC student feedback, published 2015-2021 |
| de Oliveira Lima et al. 2018 | S10 | 29 | 2013-2018 | SM | Conference | CompSci; Math | Studies using opinion mining on online hotel reviews, published 2013-2017 |
| Elnagar et al. 2021 | S11 | 60 | 2012-2020 | SLR | Conference | - | Studies using ML for Arabic SA, published 2012-2020 |
| Ghallab et al. 2020 | S12 | 108 | 2013-2018 | SLR | Journal | CompSci; Eng | Studies using Arabic SA, published 2013-2018 |
| Hajiali 2020 | S13 | 23 | 2014-2019 | SLR | Journal | CompSci; Math | Studies using big data and SA, published 2005-2019 |
| Handayani et al. 2018 | S14 | 10 | 2013-2017 | SLR | Conference | - | Studies using Malay SA, published 2008-2018 |
| He et al. 2021 | S15 | 89 | 2011-2019 | SLR | Journal | Med | Studies using computational SA methods on English social media about health topics, published 2010-2019 |
| Ibrahim & Salim 2013 | S16 | 65 | 2009-2013 | SLR | Journal | CompSci; Math | Studies using SA on Twitter, published 2003-2013 |

| Study | Study ID | Primary studies | Year range of included studies | Review type | Article type | Discipline | Inclusion |
|---|---|---|---|---|---|---|---|
| Jain et al. 2021 | S17 | 68 | 2017-2020 | SLR | Journal | CompSci; Math | Studies using ML methods for consumer SA on online reviews in English, specifically about hospitality and tourism, published 2017-2020 |
| Kastrati et al. 2021 | S18 | 92 | 2015-2020 | SM | Journal | CompSci; ChemEng; Eng; MatSci; Phys & Astron | Studies using SA on student feedback for online learning, published 2015-2020 |
| Kumar & Garg 2020 | S19 | 37 | 2009-2018 | SLR | Journal | CompSci; Eng | Studies about context-based SA, published 2006-2017 |
| Kumar & Jaiswal 2020 | S20 | 60 | 2012-2018 | SLR | Journal | CompSci; Math | Studies using supervised ML or hybrid methods of SA on Twitter text data, published 2012-2018 |
| Kumar & Sharma 2017 | S21 | 194 | 2011-2017 | SLR | Journal | CompSci; DeciSci; SocSci | Studies about opinion mining, published 2011-2017 |
| Mirzaalian & Halpenny 2019 | S22 | 85 | 2008-2018 | SLR | Journal | BMA; CompSci | Studies using social media analytics in hospitality and tourism, published 2000-2018 |
| Nassif et al. 2021 | S23 | 77 | 2015-2020 | SLR | Journal | CompSci | Studies using deep learning Arabic SA, published 2000-2020 |
| Nelci et al. 2022 | S24 | 10 | 2016-2019 | SLR | Conference | CompSci; Eng | Studies using Arabic SA, published 2015-2020 |
| Obiedat et al. 2021 | S25 | 21 | 2015-2021 | SLR | Journal | CompSci; Eng; MatSci | Studies using Arabic ABSA, published 2015-2021 |
| Osorio Angel et al. 2021 | S26 | 47 | 2012-2020 | SLR | Journal | CompSci; SocSci | Studies using Spanish SA, published 2012-2019 |
| Pinto et al. 2021 | S27 | 57 | 2015-2020 | SLR | Conference | - | Studies using SA, text mining and time series to predict stock market, published 2015-2019 |
| Polisena et al. 2021 | S28 | 12 | 2015-2020 | SCR | Journal | CompSci; Eng; MatSci | Studies using SA to assess health technologies, published 2015-2020 |
| Qazi et al. 2017 | S29 | 24 | 2002-2014 | SLR | Journal | EEF; SocSci | Studies using machine learning and concept-based SA methods on different opinion types, published 2002-2014 |
| Rana et al. 2016 | S30 | 16 | 2010-2014 | SLR | Journal | CompSci; DeciSci; Eng | Studies using ABSA or topic modelling for online reviews, published 2010-2014 |
| Salah et al. 2019 | S31 | 65 | 2004-2016 | SLR | Journal | DeciSci; BMA | Studies using SA or opinion mining on social media data, published 2004-2016 |
| Shamsi & Abdallah 2022 | S32 | 38 | 2015-2021 | SLR | Conference | CompSci; Eng | Studies using Arabic SA, published 2015-2021 |
| Sharma et al. 2020 | S33 | 10 | 2013-2018 | SCR | Journal | NeuroSci; PTP | Studies using SA on social media posts regarding pharmacotherapy, published 2002-2019 |
| Shayaa et al. 2018 | S34 | 58 | 2008-2016 | SLR | Journal | CompSci; Eng; MatSci | Studies using SA or opinion mining, published 2001-2016 |
| Skoric et al. 2020 | S35 | 74 | 2007-2019 | MA | Journal | CompSci | Studies using social media data to predict offline political behavior, published 2007-2018. |
| Tho et al. 2020 | S36 | 12 | 2016-2020 | SLR | Conference | - | Studies using ML SA on code-mixed data, published 2016-2020 |

| Study | Study ID | Primary studies | Year range of included studies | Review type | Article type | Discipline | Inclusion |
|---|---|---|---|---|---|---|---|
| Vencovsky 2020 | S37 | 13 | 2008-2019 | SLR | Conference | BMA; DeciSci; CompSci; Eng; Math | Studies using text mining on service quality, published 2008-2019 |
| Zunic et al. 2020 | S38 | 86 | 2011-2019 | SLR | Journal | HProf; Med | Studies using automatic SA methods on personal experiences related to health in text format, published 2000-2019 |

Note. SLR = systematic literature review; SM = systematic mapping study; SCR = scoping review; MA = meta-analysis; Discipline = the main disciplines of the journal/conference based on subject areas listed in Scopus, when available; AgriBioSci = Agricultural and Biological Sciences; A&H = Arts and Humanities; BMA = Business, Management and Accounting; CompSci = Computer Science; ChemEng = Chemical Engineering; DeciSci = Decision Sciences; EEF = Economics, Econometrics and Finance; Eng = Engineering; EnvirSci = Environmental Science; Hprof = Health Professions; Math = Mathematics; MatSci = Materials Science; Med = Medicine; NeuroSci = Neuroscience; Phys & Astron = Physics and Astronomy; PTP = Pharmacology, Toxicology and Pharmaceutics; SocSci = Social Science; Inclusion = key inclusion criteria for each review.

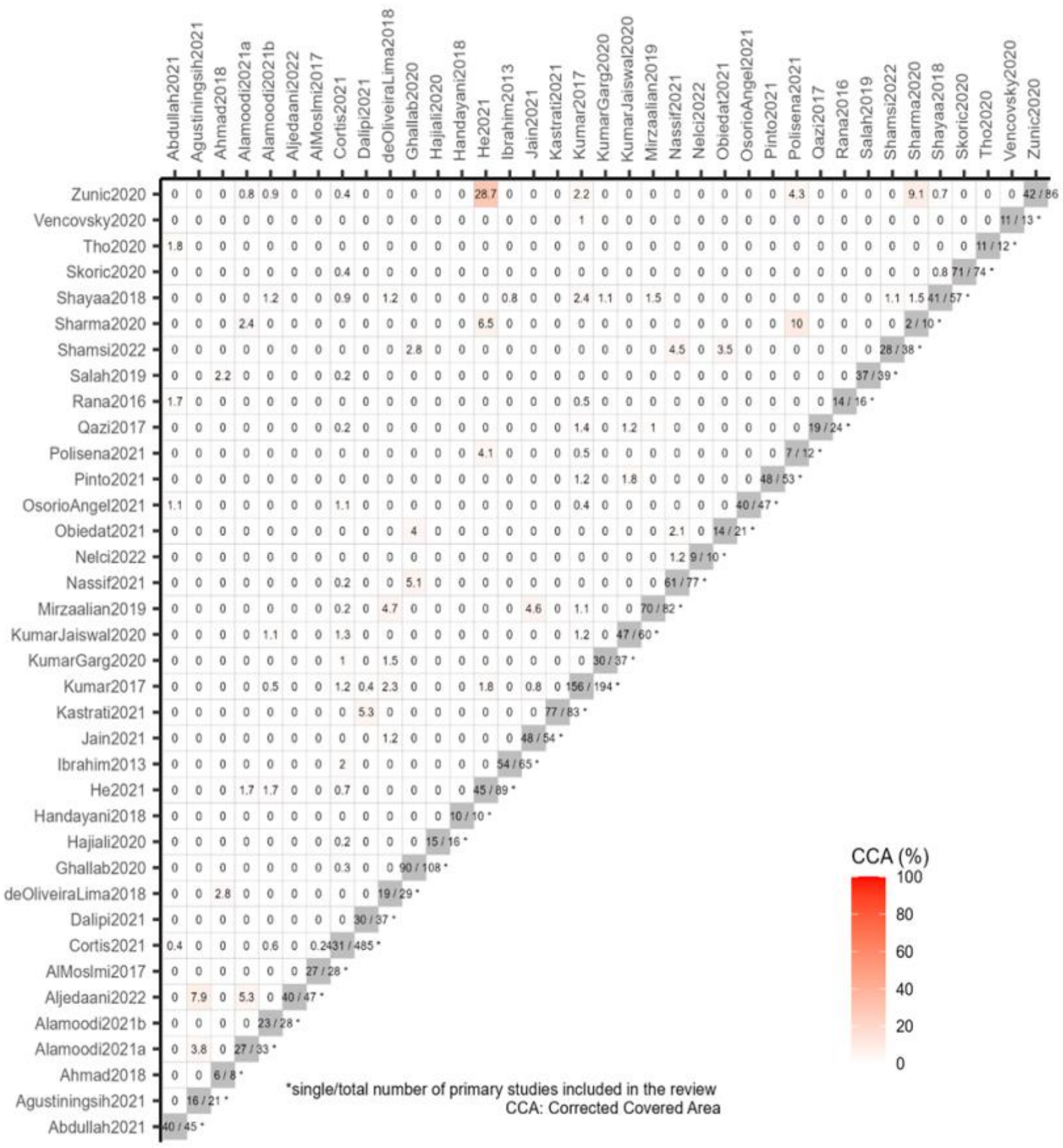

Figure 3: The pairwise corrected covered area (CCA) of the reviews

### 3.2 Specific sentiment analysis methods

SA methods were summarized and compared in varied levels of detail in the selected reviews. Furthermore, most reviews focused on a particular domain or method/s, such as SA in health or SA using deep learning methods. Table 3 summarizes the foci of the reviews, along with the most common sources, languages and modalities of the data studied.

Table 4 summarizes the most common characteristics of the SA methods, such as the pre-processing steps and classifiers used. Due to the variability in the level of detail provided, and with the aim to provide a concise overview of the key characteristics in SA methods (e.g., data sources, classifiers), we summarized the top 3 most common characteristics within each review. Similar choices for summarizing the diverse field of SA have been made in previous literature syntheses, for instance, Cortis and Davis (2021) report the top 6 most used lexicons, stating that 55 unique lexicons across studies were used, and an additional 19 studies created their own lexicons. We opted for summarizing the top 3 characteristics (e.g., data sources, classifiers), rather than the top 6, top 10 etc., due to a large variety of methods and the small number of primary studies in some of the reviews, as this would result in listing methods that were only used in a single or a handful of studies. These summaries were created to the best of our understanding, given the varying level of detail provided in the reviews. Finally, Table 5 contains brief summaries of the applications and outcomes of SA, when available. Further, see Appendix B for definitions of the technical terms used in the following summaries.

The key insights from Table 3 reveal that the reviews gauge SA literature from a variety of perspectives; the most common categories were reviews in a particular language (11 reviews), reviews of specific SA methods (10 reviews) and reviews in the health domain (8 reviews). The most common data sources in the reviews, quantified as those occurring in the top 3 sources within each review, were Twitter (22 reviews), Facebook (11 reviews) and user reviews (5 reviews), excluding reviews with a particular data source as an inclusion criterion. The languages of the data most often in the top 3 were English (7 reviews), and Chinese (5 reviews). Notably, Arabic was an inclusion criterion for seven reviews. The modality of the data was predominantly text, when specified or implied in the reviews (6 reviews). A further seven reviews only included studies using textual data.

The key insights from Table 4 are that pre-processing steps in SA were not reported in sufficient detail in the majority of the reviews (29 reviews). When these were reported, the most frequently mentioned in the top 3 were tokenization (6 reviews) and stop word removal (4 reviews). The most common features extracted for SA were TF-IDF (6 reviews), BoW, n-grams and WE (5 reviews of each), whereas features were reported in insufficient detail in 26 reviews. The ML classifiers that were most frequently reported in the top 3 were SVM (24 reviews), NB (22 reviews) and KNN (7 reviews), while the most common deep learning methods were LSTM and CNN (6 reviews each) and Bi-LSTM (3 reviews). Finally, lexicons were often not reported in detail (27 reviews), however when reported, a variety of different lexicons, including language specific lexicons, were described. The lexicons most often in the top 3 were SentiWordNet (6 reviews), LIWC, VADER and TexBlob (3 reviews each).

The key insights from Table 5 are that SA has been applied in several different domains, and a considerable number of reviews (15 reviews) did not provide systematic summaries of the applications. The applications in reviews focusing on health-related SA were more homogenous, revolving around opinions and experiences about epidemics, health services and treatments (including different medicines and vaccines). By contrast, the applications for language-focused reviews were often not specified in detail (7/11 reviews), and the remaining reviews listed varied application areas, such as economy, social and politics, or SA method development in a given language. In the reviews of specific SA methods, SA was applied in diverse settings, most notably marketing and sales, and method assessment or development. The remaining reviews were in domains of education, finance, politics, and services and tourism, with too few reviews per domain to arrive at a meaningful summary of the applications. The outcomes of SA were either summarized as the conclusions drawn from SA in the subject area (i.e., whether public opinion on Covid-19 vaccines was generally positive or negative) or as comparisons or strengths and weaknesses of different SA methods, with 20 reviews not providing synthesis of the outcomes. Method comparisons appeared particularly fragmented, such that several reviews did not provide a synthesis of the varied approaches taken in the primary studies (see section 3.3).

Table 3: Characteristics of SA data reported in the selected reviews

| Study ID | Review focus | Data source | Modality | Language |
|---|---|---|---|---|
| S1 | Language | - | - | 1. English + others 2. English + Chinese + others [c] |
| S2 | Health | Twitter (required) | - | 1. English 2. Indonesian |
| S3 | SA methods | Twitter, User reviews, Student comments | - | - |
| S4 | Health | 1. Twitter 2. Facebook 3. Scholarly journals | - | 1. English |
| S5 | Health | 1. Twitter 2. Facebook 2. Sina Weibo | 1. Text 2. Image | 1. English 2. Arabic 3. Chinese |
| S6 | Health | 1. Twitter 2. Online survey 3. Reddit | - | 1. English 2. Chinese, Japanese, Hinglish (Hindi English), Turkish |
| S7 | SA methods | 1. Amazon reviews | - | 1. English 2. Chinese 3. Norwegian |
| S8 | SA methods | 1. Twitter 2. Sina Weibo 3. Facebook | 1. Text 2. Text & image | 1. English 2. Chinese 3. Spanish |
| S9 | Education | 1. Twitter | 1. Text | - |
| S10 | Services & tourism | - | Text only (required) | - |
| S11 | Language | 1. Twitter 2. Text corpus 3. User reviews | - | Arabic (required) |
| S12 | Language | 1. Twitter 2. User reviews 3. Facebook | - | Arabic (required) |
| S13 | SA methods | - | Text only (required) | - |
| S14 | Language | 1. Twitter 2. Facebook | - | Malay (required) |
| S15 | Health | 1. Twitter 2. Health-specific online communities 3. Facebook | Text only (required) | English (required) |
| S16 | Language | Twitter (required) | - | Arabic (required) |
| S17 | Services & tourism | 1. Tripadvisor.com 2. Yelp.com 3. Online surveys | 1. Text | English (required) |
| S18 | Education | 1. Surveys & questionnaires 2. MOOC platforms 3. Social media & blogs and forums | Text only (required) | English, Chinese |
| S19 | SA methods | 1. Twitter 2. Imdb 3. Amazon | - | - |
| S20 | SA methods | Twitter (required) | Text only (required) | English (required) |
| S21 | Political | 1. Twitter [a] | Text only (required) | - |
| S22 | Services & tourism | 1. Tripadvisor and Daodao.com 2. Twitter and Sina Weibo 3. Flickr [b] | - | - |

| Study ID | Review focus | Data source | Modality | Language |
|---|---|---|---|---|
| S23 | Language | 1. Twitter 2. User reviews 3. Facebook | - | Arabic (required) |
| S24 | Language | - | - | Arabic (required) |
| S25 | Language | 1. Twitter 2. Facebook 3. Forums, Youtube, Websites | - | Arabic (required) |
| S26 | Language | 1. Twitter | 1. Text | Spanish (required) |
| S27 | Finance | Twitter, News | - | - |
| S28 | Health | 1. Twitter 2. Blogs and forums 3. Youtube, Facebook, Sina Weibo | - | - |
| S29 | SA methods | - | - | - |
| S30 | SA methods | 1. User reviews | - | 1. English 2. Chinese/ 2. multi-language |
| S31 | SA methods | - | - | - |
| S32 | Language | 1. Twitter 2. Websites (blogs, book reviews etc.) 3. Facebook | - | Arabic (required) |
| S33 | Health | 1. Twitter 2. Health-related forums | - | - |
| S34 | SA methods | 1. Twitter 2. Movie reviews 3. Amazon | - | - |
| S35 | Political | Twitter, Facebook, Forums, Blogs, Youtube | - | - |
| S36 | Language | 1. Twitter | - | 1. English-Hindi 2. English-Bengali 3. English-Telugu |
| S37 | Services & tourism | 1. User reviews 2. Twitter 3. Questionnaires | 1. Text | - |
| S38 | Health | 1. Twitter 2. Health-specific online communities 3. Facebook 3. Youtube | Text only (required) | English (required) |

Note. 1./2. /3. = the most/ the second/ the third most common data source, modality or language; when these are listed without the ranking numbers, their frequency was not clear from the review. '-' = no mention/insufficient or unsystematic description. (required) = the given data source, modality or language was an inclusion criterion for the review. [a] only a subset of 17 primary studies with focus on government intelligence, not for the full set of 194 studies. [b] for social media analytics in general, unclear what the sources for SA in particular were. [c] only multilingual primary studies were included.

Table 4: Characteristics of SA methods reported in the selected reviews

| Study ID | Pre-processing | Features | ML | DL | Transfer | Lexicons |
|---|---|---|---|---|---|---|
| S1 | 1. Machine translation 2. Tokenization 3. N-gram | - | 1. SVM 2. NB 3. KNN | - | - | - |
| S2 | 1. Stop word removal 2. Punctuation and link removal 3. Case folding / 3. Tokenization | 1. BoW / 1. TF-IDF | 1. NB 2. SVM 3. RF | 1. Bi-LSTM 2. LSTM / 2. CNN | 1. BERT | - |
| S3 | - | - | SVM (required) | - | - | - |
| S4 | - | - | - | - | - | - |
| S5 | - | - | - | - | - | - |
| S6 | - | 1. BoW / 1. TF-IDF 2. Word2Vec | 1. SVM / 1. RF 2. NB 3. KNN / 3. LR | 1. BERT 2. Bi-LSTM | - | 1. VADER 2. TextBlob 3. LIWC / 3. AmazonComprehend |

| Study ID | Pre-processing | Features | ML | DL | Transfer | Lexicons |
|---|---|---|---|---|---|---|
| S7 | - | - | - | - | - | - |
| S8 | Tokenization, Stemming, Lemmatization, NER, Dictionaries for stop words, acronyms and slang words | WE | 1. NB 2. SVM 3. LR | 1. LSTM 2. CNN 3. RNN | - | 1. SentiWordNet 2. Hu & Liu 3. AFINN/ 3. SentiStrength |
| S9 | - | - | 1. NN 2. NB 3. SVM | - | - | 1. VADER 2. TextBlob 3. SentiWordNet |
| S10 | - | - | 1. SVM 2. NB 3. LDA | - | - | 1. SentiWordNet |
| S11 | - | - | 1. SVM 2. NB 3. KNN | 1. LSTM 2. CNN | - | - |
| S12 | 1. Stemming 2. Normalization 3. Tokenization | 1. N-grams 2. TF-IDF 3. Information gain / 3. WE | 1. SVM 2. NB 3. KNN | - | - | - |
| S13 | - | - | - | - | - | - |
| S14 | - | - | 1. KNN 2. SVM / 2. NB | - | - | 1. Opinion corpus for Malay |
| S15 | - | 1. BoW 2. WE/ 2. Linguistic features (e.g., PoS and post length) | 1. SVM / 1. NB 2. LR 3. AdaBoost | - | - | 1. LIWC 2. SentimentStrength 3. LabMT |
| S16 | - | 1. N-grams 2. Combined features 3. TF-IDF / 3. BoW | 1. SVM 2. Combined classifiers 3. NB | - | - | - |
| S17 | - | - | 1. Regression 2. SVM 3. NB / 3. LR | - | - | - |
| S18 | Tokenization, pos, Normalization, Text cleaning | - | 1. NB 2. SVM 3. DT / 3. NN | - | - | 1. VADER 2. SentiWordNet 3. TextBlob / 3. Semantria |
| S19 | - | - | 1. SVM 2. LB including hybrid 3. NB | - | - | - |
| S20 | - | - | 1. SVM 2. NB 3. Ensemble methods | - | - | - |
| S21 | - | - | - | - | - | - |
| S22 | - | - | - | - | - | - |
| S23 | - | 1. WE | - | 1. CNN 2. LSTM 3. Bi-LSTM | - | - |
| S24 | - | - | - | - | - | - |
| S25 | 1. Tokenization / 1. Normalization 2. Stop word removal 3. Stemming | 1. PoS 2. N-grams 3. NER / 3. WE | 1. SVM 2. RNN 3. CNN / 3. NB / 3. KNN / 3. DT | - | - | 1. Arabic Sentiment Lexicon 2. SentiWordNet / 2. ABRL lexicon |
| S26 | 1. Tokenization 2. Pos 3. Stop word removal | 1. N-grams 2. TF-IDF 3. BoW | 1. SVM 2. Multinomial NB 3. LR | 1. CNN 2. LSTM (alone or in combination) | - | 1. LIWC 1. SOL/eSOL/iSOL |

| Study ID | Pre-processing | Features | ML | DL | Transfer | Lexicons |
|---|---|---|---|---|---|---|
| S27 | - | - | 1. SVM 2. DL 3. NN | - | - | - |
| S28 | - | - | - | - | - | - |
| S29 | - | - | 1. SVM 2. NB 3. Maximum entropy / 3. Unsupervised information extraction / 3. Relaxation labelling | - | - | - |
| S30 | - | - | 1. LDA 2. Probabilistic latent semantic analysis | - | - | - |
| S31 | - | 1. PoS 2. Frequency / 2. Syntax 3. Negation | 1. Dictionaries 2. SVM 3. NB | - | - | - |
| S32 | - | - | 1. SVM 2. NB 3. KNN | 1. CNN 2. LSTM | | - |
| S33 | Tokenization | - | 1. SVM 2. AdaBoost | - | - | 1. Hu & Liu’s opinion lexicon |
| S34 | - | - | - | - | - | - |
| S35 | - | - | - | - | - | - |
| S36 | 1. Tokenization 2. Emoticon processing / 2. URL removal 3. Stop word removal | 1. N-grams | 1. SVM 2. LR / 2. NB | - | - | - |
| S37 | - | 1. TF-IDF / 1. Absolute term frequency | - | - | - | - |
| S38 | - | - | 1. SVM 2. NB 3. DT | - | - | 1. SentiWordNet 2. Opinion lexicon 3. WordNet-Affect / 3. Multi-Perspective Question Answering |

Note. 1./2. /3. = the most/ the second/ the third most common characteristic of SA method; when these are listed without the ranking numbers, their frequency was not clear from the review. ‘-‘ = no mention/insufficient or unsystematic description. (required) = the given SA method was an inclusion criterion for the review.

Table 5: Applications and outcomes of SA reported in the selected reviews

| Study ID | Application | Outcomes |
|---|---|---|
| S1 | - | - |
| S2 | Classifying public opinion on covid-19 vaccines using twitter data | 76% of studies report higher positive than negative sentiment |
| S3 | - | Accuracy of results depends on multiple factors (e.g., pre-processing, input dataset). Multiple techniques may outperform a single technique. |
| S4 | Discovering public opinion about vaccines, understanding concerns about vaccines, etc. | - |

| Study ID | Application | Outcomes |
|---|---|---|
| S5 | Tracking epidemics, studying public response to outbreaks and to related interventions, etc. | - |
| S6 | Classifying/understanding public opinion on covid-19 vaccines | Some method comparisons: e.g., SVM and LR highly accurate, BERT outperforms ML models. Perception of covid vaccines varies by region |
| S7 | - | Pros and cons of baseline methods listed |
| S8 | About 50% of studies focused on real-world applications. Most common were politics (9%), marketing, advertising & sales (6%), technology industry (5%), finance (4%), film industry and healthcare (3% each). | Performance metrics from different studies that used the same dataset reported |
| S9 | SA effectiveness (25%), MOOC content evaluation (15%), SA via social media posts and understanding course performance and dropouts (13% each), etc. | The course content was generally rated favorably by learners across reviews |
| S10 | - | - |
| S11 | - | - |
| S12 | Most common: business and economy (37%) and social and politics domains (36%). Least common: education, health, and travel and tourism (3% each). | Model comparisons, e.g., SVM and NB compared in several studies, SVM achieved higher accuracy in 29 studies, NB had higher accuracy in 22 studies |
| S13 | - | Pros and cons: Centralized methods have high precision, are scalable, and have better performance. Distributed approaches achieve low response time, excellent scalability, and better execution time, but in some cases, low accuracy. |
| S14 | SA method development or enhancement in Malay | - |
| S15 | - | - |
| S16 | - | - |
| S17 | SA, fake reviews detection and predictive recommendation | - |
| S18 | Analyzing opinions about teachers (81%), more general opinions about teachers, courses and institutions (13%) and opinions about institutions (6%). | Comparing a sub-set of 19 high-quality papers, SA performance has mostly improved over time, due to the recent advancements (deep learning models and NLP representation techniques). |
| S19 | - | Context-based SA achieved on average 8-9% higher accuracy than SA methods not considering context and nearly 80% of studies showed the benefit of context compared to traditional SA |
| S20 | SA method improvements, assisting in decision making, enhancing intelligent analytics and personalized web experience | - |
| S21 | Market intelligence (41%), smart society services (22%), sub-component technology (13%), information and security analysis (11%), government intelligence (8%), business intelligence (4%) | - |
| S22 | Opinion mining (57/85), travel patterns (19/85), method performance testing and visitation prediction (6/85 each), where SA was barely used in the travel patterns and method performance testing | - |
| S23 | Most common: social media (59%), health services (10%), news and media (9%) | Deep learning methods tend to outperform classical classifiers |
| S24 | - | - |

| Study ID | Application | Outcomes |
|---|---|---|
| S25 | Most common domains of the datasets were hotel reviews (43%) and book reviews (19%) | Studies using ML had better accuracy than studies with LB approaches, but LB studies had better F-score than ML studies |
| S26 | - | ML typically outperforms LB approaches for specific domains, but LB performs better in open domains (especially in emotion detection). DL approaches more adaptable to other domains, compared to ML |
| S27 | Predicting stock market behavior | - |
| S28 | Assessment of patient opinions on different treatments, including vaccinations and drug therapies | - |
| S29 | - | - |
| S30 | Most common domains of the datasets were product, hotel and restaurant reviews | Due to diversity of the approaches in the review, comparison of similar techniques, datasets and languages was performed |
| S31 | Identifying user behaviors in social networks, developing marketing strategies by assessing public attitudes toward products and services, analyzing opinions on news websites, etc. | General pros and cons of ML and LB methods listed: ML: more suitable for specific domains, affected by linguistic variations problems; LB: no training required, lacks context or domain-based classification ability |
| S32 | - | - |
| S33 | Classifying opinions about a specific medication (70%), identifying adverse drug reactions (30%), analyzing sentiment dynamics in cancer forums (30%). | - |
| S34 | Most common: method assessment (28%) customer reviews (16%), marketing & sales (12%), Identifying/locating/forecasting (9%) | - |
| S35 | Using social media data to predict or gauge political behavior | ML-based SA tends to produce more precise predictions than LB SA, but typically explains less variance. Combination of ML and structural features of social network produces the most accurate predictions out of all individual approaches compared. Blogs were the best source of data for predicting political outcomes. |
| S36 | - | SVM had the highest performance (33%), followed by LR and NB (17% each) |
| S37 | - | - |
| S38 | Studying patient online communities around different health conditions, experiences of health care services (patient reviews), common obstacles for health-interventions | SVM outperformed NB (6 studies) and RF (3 studies), although SVM was also outperformed by NB (2 studies), maximum entropy and DT (1 study each). |

Note. Percentages refer to the percentage of primary studies in each review. ‘-‘ = no mention/insufficient or unsystematic description.

### 3.3 Challenges discussed in the reviews

Diverse discussion points relating to challenges and areas of improvement in future studies on SA were discussed in our selected reviews. Here we outline the most discussed issues. Firstly, the characteristics of natural language data, such as slang, negation, and spelling errors were recognized as a challenge in 12 reviews. Particularly, sarcasm and irony were considered an issue for SA systems (6 reviews). The general need for SA resources was also voiced in several reviews, such as the need for larger and publicly available datasets (6 reviews) and lexicons for specific languages and domains (5 reviews). Another common theme discussed was the need for more studies and resources in languages other than English (10 reviews), including research efforts on SA in multilingual or code-switched (i.e., alternating between two or more languages in a conversation) settings.

Importantly, 15 reviews touched the topic of difficulties in SA method comparisons, either by acknowledging factors that influence method performance and how these factors vary across studies, by noting that performance metrics were reported rarely or in varied ways, or by suggesting solutions for improving method comparison in the future. More specifically, the need for standardization in SA research was discussed (13 reviews), particularly in terms of datasets used for evaluating SA methods (6 reviews). Other concerns relating to standardization focused on reporting (6 reviews), especially regarding performance metrics – several primary studies did not report these at all, or the variability of the metrics used across studies made quantitative comparisons between SA systems difficult. These commonly recognized issues in method comparisons and standardization may explain why outcomes of SA studies regarding methods were rarely reported or summarized in the included reviews.

### 3.4 Quality of the selected reviews

We assessed the methodological and reporting quality of the selected reviews against our quality assessment framework (Table 1). We also inspected which guidelines are commonly followed in conducting SA literature reviews, and whether the use of guidelines was related to the quality of the reviews. Based on our quality assessment of the selected reviews (Table 6, scoring for each individual review is available on OSF, file Quality_Assessment.xlsx, here), the methodological rigor in synthesizing SA literature is limited. This is seen most acutely in the reporting validity of screening and data extraction, the quality assessment of primary studies, in the critical assessment of the review itself (i.e., limitations), sampling, and pre-registration of the review. These methodological steps were either not conducted or reported at all, or they were not reported in sufficient detail in over 60% of the reviews. Conversely, research objectives, inclusion/exclusion criteria and search terms were generally reported adequately, in over 65% of the reviews. Below, we outline the most critical areas of the review quality and discuss these further in section 4.3.

Table 6: Summary of quality assessment of the reviews

| Quality Assessment Criterion | Yes | Partly | No |
|---|---|---|---|
| Guidelines | 24 | 8 | 6 |
| Rationale | 15 | 13 | 10 |

| Quality Assessment Criterion | Yes | Partly | No |
|---|---|---|---|
| Objectives | 33 | 5 | 0 |
| Inclusion/Exclusion | 38 | - | 0 |
| Search Strategy: Study identification | 33 | 5 | 0 |
| Search Strategy: Search terms | 26 | 7 | 5 |
| Search Strategy: Search date | 17 | 2 | 19 |
| Included Studies | 25 | 3 | 10 |
| Validity of Screening: Methods | 9 | 5 | 24 |
| Validity of Screening: Inter-rater reliability | 2 | 0 | 36 |
| Validity of Data Extraction | 3 | 6 | 29 |
| Quality Assessment of the primary studies | 10 | 3 | 25 |
| Limitations | 14 | 0 | 24 |
| Sampling | 7 | 0 | 31 |
| Pre-registration | 0 | 0 | 38 |

#### **3.4.1***Guidelines followed in the reviews***.**

A variety of guidelines for conducting a review were followed in the selected studies. The most mentioned guideline (13 reviews) was that of Kitchenham (2004) and its updated version Kitchenham & Charters (2007), which outlines steps in conducting a systematic review in software engineering. A further eight reviews applied the PRISMA protocol, and three studies followed other guidelines (Tranfield et al., 2003, Denyer & Transfield, 2009, Arksey & O'Malley, 2005). We inspected visually whether using a guideline was associated with improved quality of the reviews and if so, whether the type of guideline used was important. To this end, the reviews were grouped as those using a guideline (Yes) and those that did not (Partly and No, see Table 1). The reviews following a guideline were further grouped into those following Kitchenham's guidelines and those following the PRISMA framework – the most common guidelines adopted in our sample. The quality of the reviews in each of these four groups ("Yes", "Partly and No", "Kitchenham", and "PRISMA", see Figure 4) were quantified as the proportion of reviews scored as 'yes' in our quality assessment. As seen in Figure 4, while following existing guidelines generally resulted in a higher quality of the reviews than not using guidelines, this was not completely consistent, and the quality of the reviews remained low for several items even when guidelines were followed. Furthermore, the review quality was not clearly associated with a particular review guideline followed (Kitchenham or PRISMA).

#### 3.4.2*Search strategy and included studies***.**

The majority of the reviews reported appropriate search strategies and clearly defined search terms. However, the time of the search performed was not reported clearly in 19 of the reviews. This hinders attempts to replicate these literature searches. Furthermore, identifying some of the included primary studies was not possible from 10 of the reviews, which prevents verifying, assessing or replicating these reviews in full.

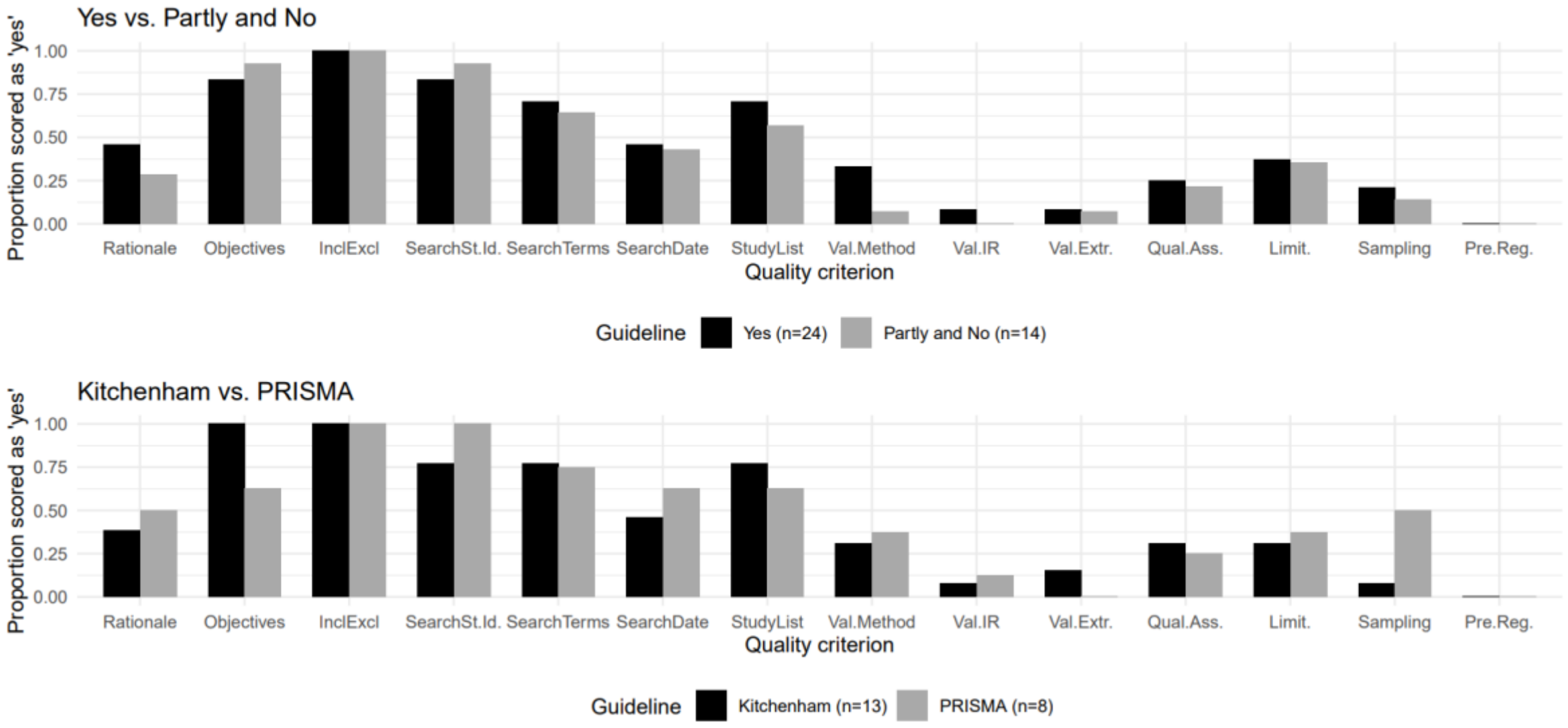

Figure 4: Proportion of reviews scored as 'yes' for each quality assessment criteria by guideline use (refer to Appendix C for a table containing the information in this bar plot)

### 3.4.3 *Validity of screening and data extraction*.

Only nine reviews clearly reported the measures taken to ensure valid study screening, and only two studies included a measure of inter-rater reliability of this process. As study screening has an element of subjectivity to it, valid screening approaches (such as calibration exercises or involvement of multiple reviewers) are essential for minimizing errors and bias in the study selection process (e.g., Higgins et al., 2023; Kitchenham & Charters, 2007; Xiao & Watson, 2019). As these procedures were generally reported inadequately or not at all in our selected reviews, it is unclear how representative the samples of the primary studies are in these reviews. Furthermore, only three reviews provide a clear description of measures taken to ensure validity of data extraction from the primary studies. As data extraction is also a process that is prone to errors and bias, it should involve calibration exercises or the involvement of multiple reviewers (e.g., Higgins et al., 2023; Kitchenham & Charters, 2007; Xiao & Watson, 2019). As this step of the review process was generally neglected, the reliability of the outcomes in the selected reviews may be compromised due to this shortcoming.

### 3.4.4 *Quality assessment and limitations*.

Quality assessment or risk of bias assessment of the selected primary studies can serve several functions, and it is recommended as the final stage of study screening or as a guide for interpreting findings (e.g., Higgins et al., 2023; Kitchenham & Charters, 2007; Xiao & Watson, 2019). Whether quality assessment was used for gaining a better picture of the potential limitations of the included studies or as a final screening for inclusion, this step of the review process was reported in only 14 reviews. This finding, coupled with the general lack of critical evaluation of the reviews themselves (reported in 10 reviews), indicates that methodological considerations do not receive enough attention in SA literature syntheses. As such, assessing the trustworthiness of the conclusions from these reviews is compromised.

#### 3.4.5 *Sampling and pre-registration*.

Only seven reviews contained some discussion of sampling, primarily in terms of data representativeness of online data. However, more critical considerations, such as risk of bias due to samples used in the primary studies were not discussed. Finally, pre-registration was not mentioned in any of the 38 reviews. As such, it is unclear how systematic the review processes were, particularly regarding whether methodological choices were pre-determined or made on an ad hoc basis. One of the benefits of pre-registering a review protocol is that it reduces the risk of bias during the review, for example in primary study inclusion (e.g., Higgins et al., 2023). Developing a protocol is also recommended in disciplines outside of medicine, although pre-registration is not considered mandatory (Kitchenham & Charters, 2007; Xiao & Watson, 2019).

### 3.5 Open science, reproducibility, and replicability

In addition to our quality assessment framework, we considered three important aspects of the reviews – the definitions of the core concept in the reviews (i.e., 'sentiment'), accessibility of the reviews and any shared material. Although these are not strictly required in some review guidelines (e.g., Kitchenham & Charters, 2007, cf. Page et al., 2021 for publicly available materials), they are nevertheless a crucial part of good scientific practices, as clear definitions of the concept of interest directly relates to construct validity of SA research (see section 4.1.1) and accessibility of the reviews and related materials promotes transparency and inclusivity of the information provided in the reviews.

#### 3.5.1 *Definitions of sentiment*.

Inspecting whether and how the core concept of sentiment was defined in the selected reviews, we found that this term was defined explicitly in only seven reviews (Kumar & Garg, 2020; Osorio Angel et al., 2021; Qazi et al., 2017). Although these definitions differed from one another, sentiment was typically described as a closely related term to opinion, attitude, emotion, feeling etc., while pointing out that these terms have subtle differences in their meanings. In 20 reviews, the meaning of sentiment was implied via description of what is analyzed in SA (e.g., Dalipi et al., 2021; Ghallab et al., 2020; Shamsi & Abdallah 2022). In these types of definitions, what was being analyzed differed across reviews, but was often equated with opinions (17 reviews), emotions (12 reviews), attitudes (9 reviews) or feelings (5 reviews). For instance, that SA is used "to analyze people's opinions and feelings about organizations, services provided, and products." (Shamsi & Abdallah 2022). The remaining 11 reviews provided uninformative, circular or no definitions for sentiment (e.g., Ahmad et al., 2018; de Oliveira Lima et al., 2018; Zunic et al., 2020). For instance, the aim of SA was described as "to automatically classify the sentiment expressed in a free text." (Zunic et al., 2020). Overall, our sample of reviews does not demonstrate consensus regarding how sentiment is defined. Explicit definitions of the term were rare, and the definitions (explicit or implied) differed from one another. However, several reviews appeared to equate sentiment with opinion, emotion, or attitude (see section 4.1.1 for discussion). These variances in definition and comprehension of what 'sentiment' actually is will have a knock-on impact regarding how we operationalize and thus measure sentiment, which is important for the inferences and conclusions made about people and data (Davidson, 2022).

#### 3.5.2 *Accessibility of the reviews and related materials*.

Out of 38 reviews, 20 were accessible from the publisher's website, and an additional four reviews were available from another source, e.g., as a pre-print. Full access to the remaining 14 articles was paywalled or subscription based. Only three reviews provided related material, such as details about the screening process or data extraction sheets. These were available upon request from authors for two reviews (Dalipi et al. 2021; Sharma et al. 2020) and provided as supplementary material for one review (He et al., 2021), which, however, was not an open access article. Overall, open science practices were widely neglected in our sample of reviews. Nearly 40% of the reviews were not freely accessible, and over 90% of the reviews lacked any supporting material that would allow full assessment and replication of the reviews.

# 4 DISCUSSION

In this overview, we synthesized the key applications, methods, outcomes, and common challenges in SA research in digital spaces, and evaluated the quality of reviews in the field. Below, we focus on considering our findings relative to the validity of SA research, expanding on commonly raised challenges in SA research and discussing the quality of literature syntheses in the SA field. This is followed by recommendation for future research for SA and NLP more broadly. We then outline limitations of this overview and end this section with concluding remarks.

## 4.1 Validity in SA research

For a method to be useful, it must be valid. For the purposes of discussing validity in SA research, we focus on construct validity, i.e., whether the method measures the phenomenon or construct it was intended to measure (e.g., Cronbach & Meehl, 1955; Bollen, 2002), concurrent validity, i.e., whether the outcome of the method correlates with that of another accepted criterion (Cronbach & Meehl, 1955), and external validity, i.e., whether the method performance can be generalized to different settings and populations (e.g., Bracht & Glass, 1968; Campbell & Stanley, 2015).

### 4.1.1*Construct validity***.**

The central concept SA systems aim to detect and classify is sentiment, which is not directly measurable (Bollen, 2002), but which should be adequately operationalized using measurable indicators of sentiment. Determining such indicators requires a clear definition of the concept of interest (Sjøberg & Bergersen, 2022). This has been largely neglected in SA literature, as no consensus on the definition of the core concepts exists. This is exemplified by our finding of missing or insufficient and diverse definitions for sentiment in the selected reviews (section 3.1.2). The lack of consensus in the terms used not only creates further incongruity in primary studies but also impacts synthesizing evidence in the field.

Munezero et al. (2014) provide an in-depth examination of concepts referring to subjectivity, such as affect, emotion, feeling, sentiment and opinion. These concepts often share synonyms in dictionary definitions, making it difficult to capture the exact differences and similarities of the phenomena these concepts refer to. Munzenero and colleagues (2014) describe how in psychological literature, more fine-grained definitions are suggested, which outline differences in the level of consciousness, duration of time and cultural influence associated with the subjective states. For example, sentiments are characterized as longer lasting and requiring an object, whereas emotions are shorter in duration and might be general rather than towards a particular object (e.g., waking up feeling sad for no clear reason). While these two subjective states are described as influenced by cultural context, opinions, by contrast, are defined as “personal interpretations of information that may or may not be emotionally charged” (Munzenero et al., 2014). Compared to these distinctions in psychological and related disciplines, more coarse measures of subjectivity appear to be adopted in NLP research, as demonstrated, for example, by commonly interchangeable use of the terms sentiment and opinion. This is why we also used both terms during our literature search, to capture as much of the relevant literature as possible.

The distinction between subjective states is not trivial when SA is used for predicting behavior, because different subjective states are associated with behavior in different ways. For example, stable attitudes are more predictive of behavior than those with less temporal stability (Doll & Ajzen, 1992; Glasman & Albarracín, 2006). Expressions of negative and positive mood online have also been shown to vary by circadian rhythm (Golder & Macy, 2011). Thus, operationalizing sentiment as any indication of subjectivity becomes problematic, as it adds noise to the analysis. For instance, if both transient expressions of emotions and longer-lasting sentiments towards a political party are equated with sentiment in SA for political forecasting, this can lead to limited predictive power and misleading results. Hence, it is critical to have agreed-upon definitions and conceptualizations of latent constructs, such as sentiment, to ensure these are measured consistently. However, akin to much of psychology, which relies heavily on natural language for theory, measurement and conceptualization, the ability to generalize findings remains much to be desired (Yarkoni, 2020). This has a domino effect when we start to look at data from people digitally and at-scale, because the way in which people use and behave across various digital devices and services changes (Davidson & Joinson, 2021) and the affordances associated with digital engagement influence behavior (Brown et al., 2022). This complicates whether and how we can capture one's sentiment consistently and accurately.

In addition to coarse definitions of sentiment, operationalizing sentiment has largely consisted of binary (positive, negative) or three-class (positive, neutral, negative) approaches. However, finer-grained classification of subjectivity is necessary for many applications. For instance, investigations on the public view of the COVID-19 vaccines would benefit from more detailed categories of emotions, such as fear and anticipation, to inform the best strategies for increasing vaccine acceptance and trust, as also discussed in Karafillakis and colleagues (2021). As such, developing more valid and effective SA systems requires a clear understanding of what exactly these systems classify, and what level of detail is needed for the systems to effectively serve their purpose.

#### 4.1.2*Concurrent validity***.**

It is customary to compare the classification of SA systems to that of human annotators for a particular dataset, which can be considered an evaluation of the concurrent validity of the system. However, the accepted external criterion for SA method evaluation (i.e., the dataset) is not well-established in the field. Instead, varied datasets are used for testing SA systems, which severely compromises comparisons of different SA methods (as frequently discussed in our selected reviews, section 3.3). Lack of standardization in method evaluations ultimately hinders progress in SA research, as the relative success of different SA systems is crucial for informing further method development. Overlooking the importance of standardized method assessment risks misplaced trust in a system's capabilities as a result of inappropriate method evaluation, and misleading conclusions if unsuitable SA systems are applied in real-world settings.

Apart from a lack of consensus on which evaluation datasets to use, the quality of these datasets also varies, which further amplifies the differences in SA method assessments. That is, some datasets used as ground-truth are more reliable than others. For instance, employing user-annotated datasets (such as star ratings of product reviews) or crowdsourced datasets where each item is labelled by a single annotator are less reliable than agreement-based, multi-annotator datasets, where only unanimous or majority annotations for each item are included (Van Atteveldt et al., 2021). Even with multiple, trained annotators, the inter-annotator agreement likely remains low, especially for more complex constructs (e.g., anxious writing rather than angry or sad or worrisome; similar to how well we can identify sarcasm if we are not part of an in-group). If we have labelled data that is not inherently reliable, any issues with these labels will feed into later training of SA systems. Further, it is important to note that cultural differences could easily come into play when labelling datasets, as the way in which emotions are described and expressed can differ between cultures (Jackson et al., 2019; Mesquita et al., 2016).

Even though several benchmark datasets for different SA tasks exist, such as the SemEval series (e.g., Pontiki et al., 2016; Rosenthal et al., 2017), Stanford Sentiment Treebank (Socher et al., 2013), and STS-Gold (Saif et al., 2013), researchers may opt to not use them for several reasons, such as their unsuitability for the specific task, language or domain of a particular SA system, or if the research targets a specific group online. This issue can be further exacerbated by the increased difficulty to obtain digital data from some online platforms, where APIs are being monetized and data access being restricted (Davidson et al., 2023). Further, it is not uncommon for digital platforms to restrict the use of data. For example, Reddit does not allow employing user-generated content for training data for machine learning, while X no longer allows inferring characteristics (e.g., political affiliation, health statuses, sexual orientation) on the individual level using their data (Davidson et al., 2023).

4.1.3*External validity***.**

This dimension of validity, relating to the generalizability of a method, is also largely neglected in SA research (Wijnhoven & Bloemen, 2014). A focus on external validity is particularly crucial when a method is applied in domains in which it was not developed or evaluated (van Atteveldt et al., 2021). For instance, investigations on general-purpose SA classifiers applied in the medical field point to high inconsistencies in method validity (He & Zheng, 2019; Weissman et al., 2019). While developing a general SA system may not even be a feasible goal due to differences in domains, cultural expressions of subjectivity (section 4.2.3) and linguistic variation in online communities (section 4.2.2), external validity of SA systems requires much more attention from the research community, so that the scope of each system is well understood and only applied in appropriate settings.

Our overview shows employment of homogenous online data sources (Twitter), language (English) and modality (text), which likely hinders the generalizability of findings from method assessments. This is because user demographics differ across social media platforms (e.g., Auxier & Anderson, 2021; Duggan & Brenner, 2013) and linguistic norms differ across online communities (Lucy & Bamman, 2021). As such, a well-performing SA system in one linguistic context may not be suitable in another (see section 4.2.2). Therefore, more comprehensive method evaluations are needed, especially regarding the types of datasets used. As SA systems are typically not tested with a variety of datasets and across domains, this likely leads to overestimation of the systems' success. Similar observations have been made about evaluation of NLP systems more generally. For instance, evaluation datasets are often "clean" (e.g., without misspellings), and as such do not resemble the type of data systems should handle in real life scenarios (Belinkov & Bisk, 2018; Rychalska et al., 2019).

### 4.2 Challenges in SA research

The most commonly mentioned challenges for SA research in our selected reviews were specific characteristics of the natural language (e.g., sarcasm, variations in spelling), need for more non-English resources (such as publicly available lexicons and datasets) and, most importantly, the need for standardization in SA research. The lack of standardization was discussed in terms of method evaluation metrics, evaluation datasets and reporting. Below, we discuss these challenges.

4.2.1*Sarcasm***.**

A common challenge recognized in SA research is detecting sentiment from utterances using figurative language, such as sarcasm (e.g., Kastrati et al., 2021; Kumar & Garg, 2020; Poria et al., 2023). This is because sarcastic comments often express the opposite sentiment to what a literal interpretation of the comment might suggest. For instance, a comment "I love this phone, it only worked for two hours!" only contains positive words, yet the sentiment it conveys is negative. Given the background knowledge that phones are expected to last longer than two hours suggests that the comment was intended to ridicule the quality of the phone rather than praise it. The human ability to correctly interpret sarcastic expressions often relies on this type of background knowledge, or world knowledge (i.e., extra-linguistic information about entities, concepts, relationships, and facts about the world). Drawing from what is known about human language processing is likely beneficial for developing more efficient sarcasm detection in NLP systems. For example, the influence of world knowledge in human language processing is demonstrated in psychological research on human sarcasm comprehension, which points to an involvement of a variety of contextual cues, such as social-cultural stereotypes regarding gender (Colston & Lee, 2004) and occupation (Katz & Pexman, 1997; Pexman & Olineck, 2002). These stereotypes are used as an aid in interpreting potentially sarcastic messages, e.g., that males are more likely to use sarcasm, or that comedians are more likely to use irony than doctors. This type of world knowledge, formed based on personal and cultural experience, is typically not available in world knowledge implemented in NLP systems, via knowledge bases, such as Wikidata (Vrandečić & Krötzsch, 2014), ConceptNet (Liu & Singh, 2004) and Atomic (Sap et al., 2019).

Apart from world knowledge, accurate interpretation of sarcasm often requires wider context than the comment or sentence of interest. For instance, determining whether the comment "he sure played well!" is sarcastic requires further context, such as "his team lost today". Sarcasm detection is particularly challenging in online content such as tweets, where contextual cues facilitating correct interpretation are often lacking due to limited length of the content.

Thus, development of more accurate NLP systems could mean 1) additional, psychologically grounded contextual/user information utilized in enhancing the reliability of sarcasm detection and 2) world knowledge representations that capture more nuanced relationships between concepts, such as those acquired in a particular cultural context.

#### 4.2.2*Linguistic variation***.**

Language use in digital spaces is diverse and varies across different online platforms and communities (Lucy & Bamman, 2021; Nguyen et al., 2016). For instance, different subreddits contain specific communication norms and specialized lingo that might be difficult to comprehend by individuals outside of the community (Robards, 2018). These community-specific language and specialized terms pose a challenge to accurate sentiment extraction if inappropriate SA resources are used. For instance, untrained or untuned language models would likely be less effective in these settings, leading to inaccurate predictions.

Apart from online language being diverse, it is also dynamic. That is, language and linguistic norms within online language communities are constantly evolving (Al-Kadi et al., 2018; Danescu-Niculescu-Mizil et al., 2013). It is therefore important to consider when the used SA resources were developed and whether the linguistic style or norms have changed since then in the intended target population. Updating resources such as lexicons and training data are important considerations in SA research and relevant aspects to report for the adequate evaluation of SA studies.

Analyzing online content is also challenging due to it containing non-standard spelling, vocabulary and syntax. Increasing the accuracy of automatic text analysis in the presence of these variations of language has been approached with normalization, that is, converting non-standard language into standard language (e.g., "imma" into "I'm going to"). However, it is not straightforward to determine where the boundaries of non-standard language lie and what exactly the standard language that online content should conform to. For instance, should "flvr" be converted in to "flavor" or "flavour"? (Eisenstein, 2013). It is also worth asking whether the process of normalization results in excluding information embedded in linguistic style that may carry meaning, such as signaling identity (Nguyen et al., 2016).

Indeed, some commonly adopted approaches in SA research may in fact hinder the ultimate goal of detecting subjective states from the content of interest. For example, stop word removal is a frequently performed pre-processing step, during which function words, such as articles (a, the), pronouns (I, he, she) and prepositions (in, on, at) are removed from the data. However, removing function words excludes potentially valuable information for the task at hand. For instance, there is evidence to suggest that the frequency of using the first person singular ("I") is a better indicator of a depressed state than analysis of negative emotion words (Chung & Pennebaker, 2011), consistent with psychological evidence linking depression/negative affect with self-focus (e.g., Pyszczynski & Greenberg, 1987; Mor & Winquist, 2002). Drawing from literature in psychological research on the purpose of function words and what they can reveal about subjective states of language users could thus be highly informative. Before function words are disregarded as noise, it would be wise to reconsider in which situations they might add to the efficiency and accuracy of NLP systems.

#### 4.2.3 *Non-English language resources*.

Our work suggests that SA has mainly been applied in English online content and, more recently, in Arabic. A focus on English is a common finding in other literature syntheses on SA (e.g., Cui et al., 2023; Ligthart et al., 2021), as well as in NLP research more generally (e.g., Blasi et al. 2022; Joshi et al., 2020). As NLP technologies occupy an increasingly significant place in today's societies, neglecting low-resource languages in this development risks silencing less central communities and hindering representation of diverse voices. For example, research in the field of international relations is largely reduced to Western- and Anglo-centric investigations, due to limited non-English language resources (Windsor, 2022). As another example, when NLP solutions provide more accurate results in resource-rich languages, the utility of technological innovation mostly serves the members of the larger language communities, creating inequality in access to economic opportunities (Weidinger et al., 2021).

In SA research, attempts to by-pass the lack of resources in certain languages has been to use machine translation, although there is some evidence suggesting that this approach decreases the quality of SA, compared to language-specific SA (e.g., Balamurali et al., 2013; Saadany & Orašan, 2020). Furthermore, the availability of language resources remains an issue in machine translation, particularly for low-resource language-pairs, hindering development of quality translation as well as the evaluation of it (see Haddow et al., 2022 for a survey). A related challenge is that concepts and their relationships are culture-specific and language-specific. As relationships between concepts differ across languages, translation equivalents may be used in different situations across languages, including concepts communicating emotion (e.g., Kollareth et al., 2018; Jackson et al., 2019). This highlights the complexity of SA across languages and suggests that mere parallel datasets for machine translation are not enough, as these mostly lack the culture-specific aspects of the given language (e.g., Akinade et al., 2023; Yao et al., 2023).

Whether increasing language diversity in NLP research is attempted by more developed machine translation or language-specific resources, these endeavors require more involvement from agents with expertise in the target languages. Some promising approaches for this are Nekoto and colleagues' (2020) participatory research in African languages, in which the participants contributed to different stages of the language resource development pipeline, such as dataset curation and benchmarking. Importantly, this project also included knowledge sharing and mentorship between international researchers and the participants with expertise in the local languages. Together these elements of the project addressed both the need for NLP technology know-how in low-resource language communities as well as the need for language and cultural expertise in each point of the language resource development.

#### 4.2.4 *Lack of standardization in method evaluation*.

Apart from a lack of consensus on the evaluation datasets that should be used in SA research (discussed in section 4.1.3), the field needs standardized procedures for method evaluation in general. This includes performance metrics used and the level of detail in reporting in SA studies (section 3.3). Inconsistent evaluation procedures and insufficient reporting of SA systems not only prevent appropriate comparison of these systems, but also the accumulation of knowledge, as replication of different studies, or further development of different systems is rendered impossible in the absence of sufficient information. This is exemplified with our finding that, for instance, pre-processing steps were not reported in sufficient detail in majority of the reviews, which may be indicative of neglecting the importance of this information, or insufficient reporting in the primary studies. These issues are prevalent in NLP field more broadly (e.g., Escartín et al., 2021; Fokkens et al., 2013; Van Miltenburg et al., 2021).

Furthermore, a deeper understanding of how different NLP systems perform and where their weaknesses lie requires method evaluation that goes beyond commonly used performance metrics (e.g., precision, recall, accuracy, F1-score, etc.). This is needed both for more realistic assessment of the systems' capabilities as well as more informative diagnostics of the systems (Ribeiro et al., 2020). For instance, detailed error analysis can offer valuable insights into when NLP systems fail, which areas of method improvement should be focused on next, and whether the errors produced by an NLP system are qualitatively similar to errors made by humans (for error analysis in SA research, see, for example, Xing et al., 2020; Zimbra et al., 2018). Apart from error analysis, more comprehensive evaluation frameworks have been developed, with focus on general linguistic capabilities of systems (Ribeiro et al., 2020) or their robustness for handling naturally occurring noise in real-life data (Rychalska et al., 2019). These types of evaluation processes should be adopted as a default approach in NLP studies, in order to achieve more reliable method assessment and to maximize the information value from method evaluations.

### 4.3 Methodological quality of literature reviews in SA

With the large and increasing volume of literature on SA, systematic literature reviews in this area are undoubtedly needed. However, the reviews should be conducted with adequate methodological rigor and transparency to avoid inaccurate representation of the accumulated knowledge in the field. The methodological rigor required in systematic reviews is essential for reducing bias (Higgins et al., 2023) and for allowing evaluation and replication of the reviews (Kitchenham & Charters, 2007; Xiao & Watson, 2019).

The methodological shortcomings we found in SA literature reviews (reported in section 3.4) raise concerns about the extent of bias introduced at different stages of the review process, such as sampling primary studies. These shortcomings also cast doubt on the reliability of the findings in the reviews, making interpretation of the conclusions from these reviews problematic. Furthermore, the issues of methodological transparency may render replication attempts impossible, thus impeding scientific progress and leading to wasted research efforts and resources (sections 3.4 and 3.5).

As researchers and policy makers rely on high quality overviews of the current state-of-the-art, unreliable systematic reviews risk misplaced research efforts and misguided policies. For instance, employing SA methods to inform government decision making without accurate knowledge of the capabilities and limitations of these methods can lead to initiatives that do not yield meaningful outcomes, thus wasting taxpayers' money and government resources. In public health, misguided policies can have direct health consequences. For example, they may lead to delayed or inappropriate responses to health crises, inadequate support service allocation, or insufficient public health measures, potentially causing harm or loss of life. These examples highlight the potential consequences of low methodological quality in SA literature reviews, and why reliable literature syntheses are paramount for evidence-based decisions on if and when to employ SA in real-life applications.

We found methodological shortcomings in several reviews, regardless of whether or which guidelines (e.g., PRISMA, Kitchenham & Charters, 2007) were followed in conducting the review. Kitchenham and colleagues (2009) also found that the quality of literature reviews in software engineering was not related to whether these reviews cited review guidelines or not. It is unclear why the quality of many reviews in these fields appears to not benefit from consulting review guidelines. This may be due to lack of expertise in review methodology or difficulties in adapting quality criteria from guidelines in different disciplines. We acknowledge that different guidelines contain different requirements for conducting a review, and that our framework for quality assessment borrows heavily from medical and health research standards (e.g.,

PRISMA). However, many of these standards of quality could be applied in SA research, either directly or with minor adaptations. It is also worth noting that the most common issues identified via application of our quality assessment framework are requirements included in review guidelines for different disciplines, such as medical and health research (Higgins et al., 2023), software engineering (Kitchemham & Charters, 2007) and planning research (Xiao & Watson, 2019).

Different publication outlets also differ in the requirements for methodological rigor, which also adds to the variability in the quality of published reviews. Some obstacles in full reporting of reviews may stem from journals' restrictions in word count or supplementary materials (Page & Moher, 2017). In fields like SA research, the issue of review quality is likely also related to the lack of widely accepted standards for reviews. Even in the case of the PRISMA reporting standards, which is widely adopted in the health and medical fields, the adherence to the reporting standards was still found suboptimal in MEDLINE publications (Page & Moher, 2017). To remedy the situation, Page and Moher suggested up-dated and clearer guidelines as well as more intense endorsement of the PRISMA statement by journals. Similarly, Stevens and colleagues (2014) point out that how journals endorse the use of guidelines is unknown, and that one possibility for ensuring adherence to guidelines would be to require reviewers to check this during peer-review process. Review and reporting guidelines have a longer history and widespread acceptance in health and medical fields, yet adherence to these guidelines still has room for improvement. Research in SA clearly has even further to go, but some of the ideas outlined above for improving adherence to review guidelines could also be implemented in the SA context, such as stricter methodological requirements evaluated as part of the peer-review process.

### 4.4 Future directions

We have discussed SA in digital spaces particularly regarding insufficient focus on different types of validity of SA research, specific challenges in SA and limited methodological and reporting quality of literature syntheses on SA.

Future work in SA should draw more heavily from psychology and related fields for clearer and more suitable definitions and operationalization of subjectivity terms of interest. Several challenges in SA due to the nature of natural language should also be re-considered from a more psychologically grounded perspective – for instance, what information is lost when normalization and stop word removal are applied to data, or how sarcasm detection could be improved. Approaching SA problems from a multidisciplinary perspective would mean not only that development of SA systems should draw from psychological research, but also that psychological research should target questions relevant to SA system development. These joint efforts would at best lead to advancement of the goals in each field. We further urge researchers to consider the variability in online data, both in terms of platform, group, and cultural differences, and in terms of change over time. At the very least, this should involve documenting these aspects of data, to allow more informed decisions regarding the appropriateness of their use in different contexts.

There is a need for commonly accepted benchmark datasets, which should be from varied sources and representative of the real-world online data (i.e., data from different online platforms and language communities, which contain misspellings and other commonly occurring characteristics of natural language in digital spaces). If suitable evaluation sets are not available, then appropriate, multi-rater annotated datasets should be created and made publicly available to allow further comparisons and thus aid moving the field forward rather than creating further fragmentation in the field.

Research in SA is plagued with non-standard and often shallow evaluations of the SA systems. For the field to move towards better understanding of the best performing and appropriate SA methods, far more focus on comprehensive and informative testing of SA systems is needed. As some examples, this could include detailed error analysis as a part of any SA study, as well as stricter requirements for full reporting and open data/code, endorsed by publication outlets.

At present, as no broadly recognized guidelines exist specifically for SA research, or NLP research more generally, development of such a guideline is paramount for improving the methodological and reporting quality in these fields. This type of guideline should include items relevant particularly for NLP work, such as detailed requirements for how to report datasets, pre-processing procedures, characteristics of the algorithms and ground truth. In the absence of a comprehensive guideline, a starting point for these improvements would be to follow existing recommendations for a dataset (Gebru et al., 2021) and model documentation (Mitchell et al., 2019). Guidelines for literature reviews should also take the specific characteristics of the field into account, to provide clear and easily applied methodology for researchers that may not be experts in literature synthesis. Another avenue for improving the quality of reviews (and primary studies) in NLP research is stricter methodological requirements as part of the peer-review process, as suggested for health research by Stevens and colleagues (2014).

### 4.5 Limitations

Systematic reviews arise when there is a need or a certain amount of evidence that warrants performing a review on a particular topic. As our overview only included systematic reviews, the topics covered here are limited to the areas of SA that have been reviewed thus far. As such, we cannot provide a complete picture of SA, particularly regarding application areas, for which we recommend the reader to consult recent reviews with particular focus on application areas (e.g., Cui et al., 2023; Kumar & Sharma, 2017; Mäntylä et al., 2018). Similarly, this overview does not represent all tasks in SA, which, as noted by Poria et al. (2023) include most of the core problems tackled by NLP in general, such as fake review detection, negation processing and domain adaptation, to name a few.

The balance between inclusion of high-quality reviews and synthesizing a representative sample of the literature is often difficult to find. This was also the case in the current overview, where our criteria for the quality of the reviews occasionally resulted in difficult decisions regarding the inclusion of reviews – some otherwise interesting works were excluded due to this and may be worth considering in broader overviews in this area. Some such reviews are Adak and colleagues' (2022) review focusing on deep learning and explainable AI used for SA on customer reviews, Kumar and colleagues' (2021) review in text mining, which included automated analyses to investigate the application domains and thematic foci of the reviews, and the review by Genc-Nayebi and Abran (2017) on text mining in mobile app store reviews.

Similarly, our focus on SA in online spaces resulted in exclusion of reviews or surveys that may still be relevant (e.g., Alshuwaier et al., 2022; Alyami et al., 2022; Mehraliyev et al., 2022), especially given that a large portion of SA research today is conducted using online content (Ligthart et al., 2021; Palanivinayagam et al., 2023), which may lead to authors not explicitly mentioning online data in the article's title or abstract.

### 4.6 Conclusion

Sentiment analysis research is a dynamic field, characterized by its diverse methodologies and wide-ranging application domains, with great promise for harnessing the rich information available in digital environments. Before we can fully embrace the potential of sentiment analysis, however, various challenges need to be addressed. These include thorough evaluation and comparison of methods, careful consideration of the scope of the methods (such as the appropriate domain and populations the methods are applied to) and commitment to sound methodological procedures and transparent reporting in both primary studies and literature reviews. These improvements should facilitate progress in the field and generate SA methodology that can be used in real-world settings with confidence and accountability.

## ACKNOWLEDGMENTS

This work has been funded by the UK Government awarded to BID & JH. The funders had no role in study design, data collection and analysis, decision to publish, or preparation of the manuscript.

## Electronic Supplementary material (p. 40-44)

## Appendix A

Appendix A Table A: Review studies included in the overview of reviews

| Study ID | Review Study |
|---|---|
| S1 | Abdullah, N. A. S., & Rusli, N. I. A. (2021). Multilingual Sentiment Analysis: A Systematic Literature Review. Pertanika Journal of Science & Technology, 29(1). |
| S2 | Agustiningsih, K. K., Utami, E., & Al Fatta, H. (2021, November). Sentiment analysis of COVID-19 vaccine on Twitter social media: systematic literature review. In 2021 IEEE 5th International Conference on Information Technology, Information Systems and Electrical Engineering (ICITISEE) (pp. 121-126). IEEE. |
| S3 | Ahmad, M., Aftab, S., Bashir, M. S., & Hameed, N. (2018). Sentiment analysis using SVM: a systematic literature review. International Journal of Advanced Computer Science and Applications, 9(2). |
| S4 | Alamoodi, A. H., Zaidan, B. B., Al-Masawa, M., Taresh, S. M., Noman, S., Ahmaro, I. Y., ... & Salahaldin, A. (2021a). Multi-perspectives systematic review on the applications of sentiment analysis for vaccine hesitancy. Computers in Biology and Medicine, 139, 104957. |
| S5 | Alamoodi, A. H., Zaidan, B. B., Zaidan, A. A., Albahri, O. S., Mohammed, K. I., Malik, R. Q., ... & Alaa, M. (2021b). Sentiment analysis and its applications in fighting COVID-19 and infectious diseases: A systematic review. Expert systems with applications, 167, 114155. |
| S6 | Aljedaani, W., Saad, E., Rustam, F., de la Torre Díez, I., & Ashraf, I. (2022). Role of Artificial Intelligence for Analysis of COVID-19 Vaccination-Related Tweets: Opportunities, Challenges, and Future Trends. Mathematics, 10(17), 3199. |

| Study ID | Review Study |
|---|---|
| S7 | Al-Moslmi, T., Omar, N., Abdullah, S., & Albared, M. (2017). Approaches to cross-domain sentiment analysis: A systematic literature review. IEEE Access, 5, 16173-16192. |
| S8 | Cortis, K., & Davis, B. (2021). Over a decade of social opinion mining: a systematic review. Artificial intelligence review, 54(7), 4873-4965. |
| S9 | Dalipi, F., Zdravkova, K., & Ahlgren, F. (2021). Sentiment analysis of students' feedback in MOOCs: A systematic literature review. Frontiers in Artificial Intelligence, 4, 728708. |
| S10 | de Oliveira Lima, T., Colaco Junior, M., & Nunes, M. A. S. (2018). Mining on line general opinions about sustainability of hotels: a systematic literature mapping. In Computational Science and Its Applications–ICCSA 2018: 18th International Conference, Melbourne, VIC, Australia, July 2-5, 2018, Proceedings, Part I 18 (pp. 558-574). Springer International Publishing. |
| S11 | Elnagar, A., Yagi, S., Nassif, A. B., Shahin, I., & Salloum, S. A. (2021). Sentiment analysis in dialectal Arabic: a systematic review. Advanced Machine Learning Technologies and Applications: Proceedings of AMLTA 2021, 407-417. |
| S12 | Ghallab, A., Mohsen, A., & Ali, Y. (2020). Arabic sentiment analysis: A systematic literature review. Applied Computational Intelligence and Soft Computing, 2020, 1-21. |
| S13 | Hajiali, M. (2020). Big data and sentiment analysis: A comprehensive and systematic literature review. Concurrency and Computation: Practice and Experience, 32(14), e5671. |
| S14 | Handayani, D., Bakar, N. S. A. A., Yaacob, H., & Abuzaraida, M. A. (2018, July). Sentiment analysis for Malay language: Systematic literature review. In 2018 International Conference on Information and Communication Technology for the Muslim World (ICT4M) (pp. 305-310). IEEE. |
| S15 | He, L., Yin, T., Hu, Z., Chen, Y., Hanauer, D. A., & Zheng, K. (2021). Developing a standardized protocol for computational sentiment analysis research using health-related social media data. Journal of the American Medical Informatics Association, 28(6), 1125-1134. |
| S16 | Ibrahim, M. A., & Salim, N. (2013). Opinion analysis for twitter and arabic tweets: A systematic literature review. Journal of Theoretical & Applied Information Technology, 56(3). |
| S17 | Jain, P. K., Pamula, R., & Srivastava, G. (2021). A systematic literature review on machine learning applications for consumer sentiment analysis using online reviews. Computer science review, 41, 100413. |
| S18 | Kastrati, Z., Dalipi, F., Imran, A. S., Pireva Nuci, K., & Wani, M. A. (2021). Sentiment analysis of students' feedback with NLP and deep learning: A systematic mapping study. Applied Sciences, 11(9), 3986. |
| S19 | Kumar, A., & Garg, G. (2020). Systematic literature review on context-based sentiment analysis in social multimedia. Multimedia tools and Applications, 79, 15349-15380. |
| S20 | Kumar, A., & Jaiswal, A. (2020). Systematic literature review of sentiment analysis on Twitter using soft computing techniques. Concurrency and Computation: Practice and Experience, 32(1), e5107. |
| S21 | Kumar, A., & Sharma, A. (2017). Systematic Literature Review on Opinion Mining of Big Data for Government Intelligence. Webology, 14(2). |
| S22 | Mirzaalian, F., & Halpenny, E. (2019). Social media analytics in hospitality and tourism: A systematic literature review and future trends. Journal of Hospitality and Tourism Technology, 10(4), 764-790. |
| S23 | Nassif, A. B., Elnagar, A., Shahin, I., & Henno, S. (2021). Deep learning for Arabic subjective sentiment analysis: Challenges and research opportunities. Applied Soft Computing, 98, 106836. |
| S24 | Nelci, J., Salloum, S. A., & Shaalan, K. (2022). An overview of sentiment analysis with dialectical processing. In Proceedings of International Conference on Emerging Technologies and Intelligent Systems: ICETIS 2021 (Volume 1) (pp. 3-12). Springer International Publishing. |
| S25 | Obiedat, R., Al-Darras, D., Alzaghoul, E., & Harfoushi, O. (2021). Arabic aspect-based sentiment analysis: A systematic literature review. IEEE Access, 9, 152628-152645. |
| S26 | Osorio Angel, S., Pena Perez Negron, A., & Espinoza-Valdez, A. (2021). Systematic literature review of sentiment analysis in the Spanish language. Data Technologies and Applications, 55(4), 461-479. |
| S27 | Pinto, N., da Silva Figueiredo, L., & Garcia, A. C. (2021, May). Automatic Prediction of Stock Market Behavior Based on Time Series, Text Mining and Sentiment Analysis: A Systematic Review. In 2021 IEEE 24th International Conference on Computer Supported Cooperative Work in Design (CSCWD) (pp. 1203-1208). IEEE. |

| Study ID | Review Study |
|---|---|
| S28 | Polisena, J., Andellini, M., Salerno, P., Borsci, S., Pecchia, L., & Iadanza, E. (2021). Case Studies on the Use of Sentiment Analysis to Assess the Effectiveness and Safety of Health Technologies: A Scoping Review. IEEE Access, 9, 66043-66051. |
| S29 | Qazi, A., Raj, R. G., Hardaker, G., & Standing, C. (2017). A systematic literature review on opinion types and sentiment analysis techniques: Tasks and challenges. Internet Research. |
| S30 | Rana, T. A., Cheah, Y. N., & Letchmunan, S. (2016). Topic Modeling in Sentiment Analysis: A Systematic Review. Journal of ICT Research & Applications, 10(1). |
| S31 | Salah, Z., Al-Ghuwairi, A. R. F., Baarah, A., Aloqaily, A., Qadoumi, B. A., Alhayek, M., & Alhijawi, B. (2019). A systematic review on opinion mining and sentiment analysis in social media. International Journal of Business Information Systems, 31(4), 530-554. |
| S32 | Al Shamsi, A. A., & Abdallah, S. (2022). A Systematic Review for Sentiment Analysis of Arabic Dialect Texts Researches. In Proceedings of International Conference on Emerging Technologies and Intelligent Systems: ICETIS 2021 Volume 2 (pp. 291-309). Springer International Publishing. |
| S33 | Sharma, C., Whittle, S., Haghighi, P. D., Burstein, F., & Keen, H. (2020). Sentiment analysis of social media posts on pharmacotherapy: A scoping review. Pharmacology Research & Perspectives, 8(5), e00640. |
| S34 | Shayaa, S., Jaafar, N. I., Bahri, S., Sulaiman, A., Wai, P. S., Chung, Y. W., ... & Al-Garadi, M. A. (2018). Sentiment analysis of big data: methods, applications, and open challenges. IEEE Access, 6, 37807-37827. |
| S35 | Skoric, M. M., Liu, J., & Jaidka, K. (2020). Electoral and public opinion forecasts with social media data: A meta-analysis. Information, 11(4), 187. |
| S36 | Tho, C., Warnars, H. L. H. S., Soewito, B., & Gaol, F. L. (2020, November). Code-Mixed Sentiment Analysis Using Machine Learning Approach–A Systematic Literature Review. In 2020 4th International Conference on Informatics and Computational Sciences (ICICoS) (pp. 1-6). IEEE. |
| S37 | Vencovský, F. (2020). Service quality evaluation using text mining: A systematic literature review. In Perspectives in Business Informatics Research: 19th International Conference on Business Informatics Research, BIR 2020, Vienna, Austria, September 21–23, 2020, Proceedings 19 (pp. 159-173). Springer International Publishing. |
| S38 | Zunic, A., Corcoran, P., & Spasic, I. (2020). Sentiment analysis in health and well-being: systematic review. JMIR medical informatics, 8(1), e16023 |

## Appendix B

General Definitions of Technical Terms Used in Studies

Appendix B Table B: Definitions of technical terms used in studies

| Abbreviation | Full version | Brief Definition |
|---|---|---|
| AdaBoost | Adaptive Boosting | AdaBoost combines multiple “weak” learners (e.g., decision trees) to create a strong classifier. By doing this, it can build a powerful ensemble model that performs well on complex classification problems. |
| Bi-LSTM | Bidirectional Long-Short Term Memory | The process of making neural networks to be able to sequence information in both directions (backwards and forwards). Hence, input flows in two directions thus making them unique from (unidirectional) LSTM networks. |
| BoW | Bag of Words | An approach to text representation used in natural language processing (NLP) and information retrieval (IR). Text (e.g., sentences, social media posts, documents) is represented as the bag (or multiset) of its words, whereby we disregard grammar and word order without compromising multiplicity. |
| BERT | Bidirectional Encoder Representations from Transformers | A deep learning model, specifically designed for sequence transduction tasks, such as language translation and text generation. BERT can consider both the left and right context of a word during training, allowing it to capture the contextual information of words more effectively than models that only consider the left or right context. |

| Abbreviation | Full version | Brief Definition |
|---|---|---|
| CNN | Convolutional Neural Network | A CNN is a type of artificial neural network, typically applied to image analysis. They consist of multiple layers, including hidden layers, where at least one of those hidden layers will be convolutional layers. |
| DL | Deep Learning | DL is a type of machine learning that uses increasing numbers of layers to extract higher-level features from the raw input. These were based on CNNs. |
| DT | Decision Tree | DTs are a type of algorithm used in supervised machine learning approaches. DTs can have a categorical/discrete target variable (e.g., colour, gender) known as a classification tree, or DTs can have a continuous target variable (e.g., price, height) known as a regression tree. They are particularly interpretable and simple to understand due to the tree-like rules they produce. |
| EC | Evolutionary Computing | EC is a set of algorithms typically associated with optimization problems. In essence, a EC algorithm will produce an initial set of solutions and it is iteratively updated. Hence, it generates a population of solutions, which are iteratively subject to natural selection (artificial selection) and mutation over time to find an optimal (or other) solution. |
| EM | Ensemble Methods | EMs use multiple machine learning algorithms to improve predictive performance in supervised problems. For example, a Random Forest is an ensemble method that generates n trees and the most common prediction is taken (in a classification setting). There are many types of ensemble methods, e.g., bagging or boosting. |
| FL | Fuzzy Logic | FL has a variety of applications. In its main form, it acts as a way to convey partial truth, where the truth of the variable is any real number between 0 and 1. |
| GB | Gradient Boosting | GB is a type of machine learning algorithm used for regression and classification tasks. It is a type of ensemble method, typically using DTs. |
| GRU | Gated Recurrent Unit | GRU is a gating mechanism in RNNs. They are similar to LSTMs, where they have forget gates (which determines what information can be transferred), however GRUs have less parameters and no output gate. |
| KNN | K Nearest Neighbours | KNN is a supervised machine learning algorithm, which can be applied to classification and regression problems. It is a type of lazy learner, whereby it only stores the training dataset verbatim, rather than having two distinct stages of training and testing. |
| LDA | Latent Dirichlet Allocation | In NLP, LDA is a generative statistical model that aims to explain a set of observations via unobserved groups. Hence, it is conceptually similar to cluster analysis. It is commonly used in NLP for topic modelling, where observations (e.g., words, sentences) are collated into documents, and the words are then attributed to topics across the documents. |
| LB | Lexicon-based | LB approaches in sentiment analysis (SA) is an approach that helps to classify text into sentiments - positive through to negative. Typically, these rely on the words themselves and they find the sentiment score of the words themselves and aggregates to the observation level (e.g., posts, sentences) |
| LR | Logistic Regression | A type of model that models the probability of an event taking place, where the log odds for the event are a linear combination of the feature variables. These have binary target variables. |
| LSTM | Long-Short Term Memory (networks) | Are a type of Recurrent Neural Network (RNN), which was developed to avoid long-term dependency problems. Hence, LSTM networks are able to remember information for long periods. |
| ML | Machine Learning | ML is a wide field that is interested in being able to extract knowledge via patterns in the raw data. There are many subdomains of this, including unsupervised learning (clustering), supervised learning, and reinforcement learning. |
| MLP | Multilayer Perceptron | MLPs are a type of feedforward artificial neural network with fully connected layers. |
| NB | Naïve Bayes | NB is a type of probabilistic classifier based on Bayes' theorem. They work well with multi-class problems and are scalable, however they require independence assumptions to be met between features. |
| NER | Named Entity Recognition | NER seeks to identify and classify named entities in unstructured text (e.g., documents, social media posts) into predefined categories (e.g., names, organisations) |

| Abbreviation | Full version | Brief Definition |
|---|---|---|
| NN | Neural Network | NN also known as artificial neural networks (ANNs) are algorithms inspired by the brain. They consist of connected nodes (artificial neurons) and their connections (synapses) transmit signals to other neurons -- thus conceptually similar to the brain. Neurons are aggregated into layers: input and output, with some having multiple layers in between (e.g., hidden layers, convolutional layers). |
| PR | Probabilistic Reasoning | PR uses probability and logic to handle uncertainty. |
| RF | Random Forest | A RF is an ensemble method for classification and regression tasks. For classification tasks, the most common class is selected (mode), and for regression trees the mean or average prediction is taken (mean). They are often provide a good performance from the outset. |
| SVM | Support Vector Machine | SVMs are a form of supervised learning, different kinds can perform both classification and regression tasks. Different forms of kernels can be used, which allows SVMs to tackle both linearly separable and non-linearly separable problems with a relatively small number of training examples. |
| T | Transformers | A transformer is a type of DL architecture increasingly used in NLP. They process text directionally, which provides more context and information about the sentence. Their in-built self-attention mechanisms allow the model to differentially weightings the significance across inputs, thus allowing insight into the relationships between words and phrases across the documents. |
| TF-IDF | Term Frequency-Inverse Document Frequency | TF-IDF is a statistic that reflects the importance of words to a document or corpus. This considers both the term frequency (TF), which would highlight very common words (e.g., the) and the inverse document frequency (IDF), which when incorporated with TF diminishes the weight of terms occurring highly often. Hence, this helps adjust words that are highly used against rarer words. |
| WE | Word Embedding | In NLP, WBs are representations of words that are used for a variety of later analyses. These representations encode the meaning of the word so words that are closer in the vector space are likely to have similar meaning. |

**Appendix C**

Table containing the information presented in Figure 4: Proportion of reviews scored as ‘yes’ for each quality assessment criteria by guideline use (Yes, Partly and No, Kitchenham, and PRISMA)

| Quality criterion | Yes (n=24) | Partly and No (n=14) | Kitchenham (n=13) | PRISMA (n=8) |
|---|---|---|---|---|
| Rationale | 0.46 | 0.29 | 0.38 | 0.50 |
| Objective | 0.83 | 0.93 | 1.00 | 0.63 |
| Inclusion/Exclusion | 1.00 | 1.00 | 1.00 | 1.00 |
| Search Strategy: Study Identification | 0.83 | 0.93 | 0.77 | 1.00 |
| Search Strategy: Search Terms | 0.71 | 0.64 | 0.77 | 0.75 |
| Search Strategy: Search dates | 0.46 | 0.43 | 0.46 | 0.63 |
| Included Studies | 0.71 | 0.57 | 0.77 | 0.63 |
| Validity of Screening: Methods | 0.33 | 0.07 | 0.31 | 0.38 |
| Validity of Screening: Inter-rater Reliability | 0.08 | 0.00 | 0.08 | 0.13 |
| Validity of Data Extraction | 0.08 | 0.07 | 0.15 | 0.00 |
| Quality Assessment of the Primary Studies | 0.25 | 0.21 | 0.31 | 0.25 |
| Limitations | 0.38 | 0.36 | 0.31 | 0.38 |
| Sampling | 0.21 | 0.14 | 0.08 | 0.50 |
| Pre-registration | 0.00 | 0.00 | 0.00 | 0.00 |